\documentclass[11pt]{article}
\usepackage[round]{natbib}
\usepackage{graphicx}
\usepackage{soul}
\usepackage{marginnote}

\newcommand{\fn}[1]{\footnote{\renewcommand{\baselinestretch}{1}\small\footnotesize #1}}

\newcommand{\para}[1]{\begin{quote}\renewcommand{\baselinestretch}{1}\normalsize\small #1 \end{quote}\renewcommand{\baselinestretch}{1.5}\small\normalsize}

\begin{document}
\title{On the Ontology of Particle Mass and Energy in Special Relativity}
\author{Forthcoming in \emph{Synthese} \\ $\textrm{ }$ \\ Kevin Coffey\thanks{Thanks to Gordon Belot, Carl Hoefer, Mario Hubert, Tim Maudlin, Trevor Teitel, and two anonymous referees from \emph{Synthese} for helpful comments and suggestions, and to audiences in Dubrovnik, Helsinki, London, Prague, San Sebastian, and Winston-Salem. I am particularly indebted to Sam Fletcher and Chip Sebens for extensive discussion and feedback, and to the John Bell Institute for the Foundations of Physics in Hvar for providing a serene and glorious environment for finishing the paper.} \\ Department of Philosophy \\ New York University Abu Dhabi}

\date{}
\maketitle

\begin{abstract}
\noindent Einstein claimed that the fundamental dynamical insight of special relativity was the equivalence of mass and energy. I disagree. Not only are mass and energy not equivalent (whatever exactly that means) but talk of such equivalence obscures the real dynamical insight of special relativity, which concerns the nature of 4-forces and interactions more generally. In this paper I present and defend a new ontology of special relativistic particle dynamics that makes this insight perspicuous and I explain how alleged cases of mass--energy conversion can be accommodated within that ontology.
\end{abstract}

\renewcommand{\baselinestretch}{1.5} \small\normalsize

\section{Introduction}

Special Relativity is widely recognized as having transformed our understanding of the relationship between energy and mass.\fn{In the words of one mathematician--historian, the energy--mass relationship brought about by special relativity constitutes ``the most striking example of unification that has been effected in the present century''. \citep[p.96]{ewhittaker58}} Over a decade after the theory's development, Einstein wrote that the special relativistic reconceptualization of the mass--energy relationship, as expressed in the equation $E=mc^2$, was the theory's most important and lasting contribution to our understanding of the physical world---not the relativity of simultaneity, time dilation, or length contraction.\fn{\citet[p.230]{aeinstein19}. See also \citet[p.73]{wrindler91}} When one reflects on the role this equation played in the development of nuclear weapons and nuclear energy---processes in which mass is apparently converted into energy---it's difficult to disagree. Mass and energy seem to bear a substantive physical connection to each other in the relativistic context that is quite unlike their classical relationship.

Although the \emph{mathematical} expressions relating mass and energy are straightforward, the precise nature of the \emph{physical} reconceptualization remains poorly understood. Are mass and energy the same physical property, or are they instead distinct but related quantities? What does it mean to say that mass is ``converted'' into energy (or energy into mass)? Does $E=mc^2$ suggest some new, non-classical ontology of mass and energy?

The slogans physicists use to characterize the mass--energy relationship provide little clarity. \citet{rdinverno92} writes that \para{[$E=mc^2$] is not just a mathematical relationship between two different quantities, namely energy and mass, but rather states that energy and mass are \textbf{equivalent concepts}. (48)\fn{See also \citet[pp.16--20]{afrench68}, \citet[p.493]{sternheimkane91}, and \citet[pp.306--307, n.13]{rtorretti83}, who equate energy and mass either conceptually or metaphysically. This view is echoed in the philosophical literature by, e.g., \citet[p.104]{jbutterfield84}, \citet[p.18]{jearman89}, and \citet[p.382]{pteller91}.}}
\citet{vfaraoni13} puts the relationship in a seemingly different way: \para{[$E=mc^2$] expresses the fact that a free particle possesses energy just because it has mass, and that a small mass can free up an enormous amount of energy because the factor $c^2$ is large in ordinary units (as demonstrated in nuclear reactions and nuclear bombs): this is the \emph{equivalence of mass and energy}. (144)\fn{\citet[pp.143--152]{thelliwell10} expresses a similar view, whereas \citet[pp.81--84]{wrindler91} seems to combine both passages. \label{phys}}

}
Do these authors agree on the nature of the alleged equivalence? Strikingly, \citet{bondispurgin87} reject any apparent equivalence between mass and energy: \para{Mass and energy are not interconvertible. They are entirely different quantities and are no more interconvertible than are mass and volume, which also happen to be related by an equation, $V=m\rho^{-1}$...The best way to appreciate Einstein's conclusion is to realise that energy has mass...[All] should be warned against believing erroneous statements that mass and energy are interconvertible, and they should be urged to avoid such terminology as `the equivalence of mass and energy'. (62--63)}
Evidently, not all of these physicists can be right.

This paper proposes a new ontology for the dynamics of special relativistic particles that is motivated by an attempt to clarify the relationship between energy and mass, as understanding that relationship is ultimately grounded in what the fundamental physical properties of particles \emph{are}. I argue that energy and mass are not equivalent: the appearance of inter-conversion is the product of an inadequate (if ubiquitous) view of the underlying dynamics.\fn{\citet{fflores05} identifies six interpretations of Einstein's equation represented in the literature on special relativity, tentatively endorsing the view that mass and energy are inequivalent physical properties that can---but need not---be converted into each other. I think that both his positive argument and his grounds for rejecting several of the competing interpretations rest on conceptual misunderstandings, but will confine my commentary to footnotes. The view developed here is not among the six interpretations Flores canvasses.} On the view developed here, the surprising and central dynamical insight of special relativity lies not in any relationship between energy and mass, as Einstein claimed, but rather in the nature of \emph{interactions} between particles mediated by 4-forces in Minkowski spacetime.\fn{\label{method}This paper is restricted to the special relativistic dynamics of (spinless) particles. One might feel that a clear understanding of energy--mass `equivalence' can't be adequately addressed independently of general relativity or broader field-theoretic considerations. See \citet[p.454, n.1]{dlehmkuhl11} for an expression of this attitude. But there are good reasons to think the energy--mass relationship can be investigated in an illuminating way in the limited context of special relativistic particle dynamics, and in fact that such a restricted context is the appropriate starting point for an inquiry into the relationship between energy and mass. First, the original association of mass with energy, articulated in \citet{aeinstein05}, draws solely upon special relativistic particle dynamics. There is thus a straightforward conceptual question about how such an equivalence is to be understood that predates any general relativistic or field-theoretic considerations. Einstein thought the identification of mass and energy was already grounded in the comparatively simple relativistic theory of point particle dynamics. Second, the philosophical challenges raised to the received view discussed below resurface in the broader context of general relativity. As noted by \citet{choefer00}, the conceptual status of energy and mass is, if anything, \emph{more} problematic in that context. It is thus good philosophical methodology to start with the simpler case in the hopes that a clear understanding of special relativistic particle dynamics might point the way towards understanding more elaborate contexts. Whether the interpretation developed here can be suitably extended to classical fields, including general relativity, is an open question.
}

Section two presents the account of mass and energy implicit in most textbook characterizations and some of the physical considerations motivating it. The central difficulty confronting this view is an apparent confusion in treating mass and energy as having the same ontologically fundamental status, a puzzle that has been discussed in \citet{mlange01, mlange02a} and is outlined in section three. Lange's own `perspectival' account of the mass--energy relationship is critiqued in sections four and five. In the process I raise several overlooked interpretive puzzles for understanding relativistic dynamics. The alternative ontology developed in section six takes the fundamental properties of particles to be encoded in geometrical features of their 4-momenta, and in subsequent sections I use those properties to propose a new account of what makes special relativity \emph{dynamically} (as opposed to \emph{kinematically}) novel---an account on which there is no deep ontological connection between energy and mass.

\section{The Received View}
\subsection{Mass as a Form of Energy}

Traditional presentations of energy and mass in special relativity generally proceed from the definition of the relativistic energy of a free particle: $$E = \gamma_u mc^2.$$ Here $m$ is the particle's mass, $c$ is the speed of light, $u$ is the particle's speed, and $\gamma_u$ is the Lorentz factor given by $$\gamma_u = \frac{1}{\sqrt{1-u^2/c^2}}.$$ In the relativistic context, unlike the classical one, a particle is evidently recognized as possessing energy simply in virtue of possessing mass, for in a frame in which a particle is stationary ($u=0$) the relativistic energy expression reduces to $E_0=mc^2$, where $E_0$ is called the \emph{rest energy}.\fn{Here `mass' is being used in the modern sense of the property that determines how a particle resists changes to its state of motion (see, e.g., \citet[p.2]{tmoore13}). \citet{inewton99} famously thought of mass differently---as in some sense a measure of a body's `quantity of matter'. \citet{mjammer97} discusses the history of this conceptual transformation, which has its origins in Euler's work in the 18th century. For a philosophical justification of this reconceptualization, see \citet{hcartwright75} and \citet{mlange01}. The quantity represented by $m$ is sometimes misleadingly called a particle's `rest mass', although I follow \citet{mlange01, mlange02a}, \citet{tmoore13}, \citet{wrindler91}, \citet{rwald84}, and others in treating it as an intrinsic, frame-independent property of a particle. That \citet{fflores05} fails to appreciate this point leads to a misunderstanding---and misplaced criticism---of Lange's view. (See note \ref{flores2} below.) Mass should be distinguished from a particle's so-called `relativistic mass', given by $m_R=\gamma_um$, for which rest mass is the special case corresponding to $u=0$. \label{flores1}} That rest energy stands in a fixed ratio to mass has suggested to many physicists that mass \emph{just is} rest energy (sometimes suggestively called \emph{mass-energy}). For example, \citet{taylorwheeler92} write that
\para{[$E_0=mc^2$] is the most famous equation in all physics. Historically, the factor $c^2$ captured the public imagination because it witnessed to the vast store of energy available in the conversion of even tiny amounts of mass to heat and radiation. The units of $mc^2$ are joules; the units of $m$ are kilograms. However, we now recognize that joules and kilograms are units different only because of historical accident. The conversion factor $c^2$, like the factor of conversion from seconds to meters or miles to feet, can today be counted as a detail of convention rather than as a deep new principle. (pp.203--206; diagrams on pp.204--205)}
This understanding, which I take to be held by many physicists, holds that mass and rest energy are ``equivalent'' in the sense that they are one and the same physical property; \emph{their terms are coreferential}. Given the interconvertability between different forms of energy, one might also say that mass is equivalent to energy \emph{in general}, and that mass and kinetic (or potential) energy are ``two forms of the same thing''.\fn{\citet[p.493]{sternheimkane91}. See also \citet[pp.34--35]{tmoore13}.} Mass and energy are interconvertible in the same sense that different forms of energy are interconvertible.

\subsection{Mass Defects and Inelastic Collisions}
\label{inelastic}
But channeling \citet{bondispurgin87} in the passage above, we don't \emph{identify} photon energy ($E$) and frequency ($\nu$), even though they, too, are related in fixed proportion as $E=\hbar\nu$, nor do we treat energy and frequency as ``two forms of the same thing''. So why interpret $E_0=mc^2$ differently? Why not say that a certain amount of energy ($E_0$) is `associated' with an object's mass?

The central reason is that there seem to be physical processes in which mass and other forms of energy are \emph{interconverted}. One oft-cited illustration concerns `mass defects' associated with radioactive decay. When a tritium (or hydrogen-3) nucleus decays in a nuclear fission reaction into a helium-3 nucleus, an electron, and a neutrino $$^3_1H \rightarrow \textrm{ } ^3_2He^{1+} + e^- + \bar{\nu}_e,$$ the mass of the tritium is found to be \emph{greater} than the sum of the masses of the individual daughter bodies. Part of the mass of the tritium has apparently been converted into the kinetic energies of the outgoing bodies, for the loss of mass corresponds to a loss of rest energy determined by $\Delta E_0=\Delta mc^2$---precisely the amount needed to account for the \emph{increased} kinetic energy.

A second example concerns inelastic particle collisions. Consider a generic inelastic collision between two particles of equal mass, as viewed from their center-of-momentum frame: \\

  \begin{center}
	\includegraphics[width=3in]{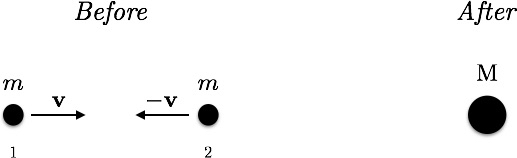}
  \end{center}

\noindent Prior to collision, the mass and energy of the system are given by $$m_{sys} = m_1 + m_2 = 2m$$ $$E_{sys} = E_1 + E_2 = 2\gamma_umc^2.$$ Given energy conservation, the final system energy is now also given by $$E_{sys} = \gamma_{u^\prime}Mc^2 = Mc^2,$$ where $u^\prime = 0$ and $\gamma_{u^\prime} = 1$. Equating both expressions for system energy $$2\gamma_umc^2 = Mc^2$$ implies that $$M = 2\gamma_um > 2m.$$ Here mass appears to be \emph{gained} in the collision process, and this gain is accompanied by a corresponding loss in the system's kinetic energy. Kinetic energy has apparently been converted into mass, for the loss in kinetic energy corresponds precisely to the gain in actual mass one would expect if mass and rest energy were two forms of the same thing.\fn{The kinetic energy ($T$) of a free particle in special relativity is given by $T=(\gamma_v - 1)mc^2$ so that $E=\gamma_vmc^2=E_0 + T$. In the inelastic collision, the kinetic energy lost is $2(\gamma_v - 1)mc^2$, which corresponds precisely to the rest energy (and hence mass) gained: $Mc^2 - 2mc^2 = 2\gamma_vmc^2 - 2mc^2 = 2(\gamma_v - 1)mc^2$. \label{kinetic}}

\section{The Problem of Lorentz Invariance}

This widespread understanding of the energy--mass relationship has been challenged by \citep{mlange01, mlange02a}, who argues that the differing ontological statuses of mass and energy in the theory undermines their alleged equivalence. Lange's argument is grounded in two related and widely-accepted features of special relativity. First, that all inertial frames are physically equivalent in the sense that the dynamical laws take the same form in all inertial frames and, on their basis, do not permit the identification of any one such frame as physically privileged or distinguished.\fn{There are subtle issues concerning how to make precise sense of this notion of equivalence and of the role of symmetries in theory interpretation in general, but that is a topic for another paper. That the physical equivalence of all inertial frames lies at the foundation of special relativity is explicit in \citet[pp.1,7,50]{wrindler91}, and Lange's invocation of the notion of Lorentz-invariance is standard in both the physics and philosophy literatures.} Second, and relatedly, that any numerical quantities representing primitive or basic physical properties are Lorentz-invariant---that is, are not frame- or observer-dependent. (This is of course not to suggest that all fundamental physical properties must be represented by scalar quantities.) Versions of both claims are also generally taken to hold for classical dynamics, although in special relativity the class of such frame-dependent (and so non-fundamental) quantities is larger and more counter-intuitive.\fn{Throughout this paper I make use of the distinction between a fundamental physical property and a derivative or non-fundamental physical property. However those metaphysical notions are made precise, it is this author's opinion that the distinction is a substantive one, and moreover one that is implicit in physical practice.
}

Consider how these interpretive commitments get applied to discussions of relativistic \emph{length} and \emph{distance}. As standardly understood, there is no fundamental fact in special relativity about how long a measuring rod is, as its length (as represented by a number) is not Lorentz-invariant. Its value changes from one frame to the next in accordance with the Lorentz transformations. There are facts about the length of the rod in different frames---there is a perfectly real and objective physical fact about its length in the inertial frame in which it's at rest (its `rest length'), say, assuming such a frame exists---but there is no frame-independent fact about \emph{the} length of the rod. Length is not an absolute property of the rod, and thus not a candidate for being fundamental. None of this is to suggest that we ought not to be interested in such derivative frame-dependent physical facts and properties. Indeed, such features of the world are often the most readily accessible and convenient with which to work, given the modes of description that come naturally to us. But this is all the more reason why we must be cautious in making judgments about the theory's fundamental ontology.\fn{The interpretive constraint that the mathematical objects and equations characterizing a theory's fundamental ontology satisfy certain sorts of symmetry requirements---in this case, requirements associated with Lorentz transformations---is not without its philosophical puzzles. See, e.g., \citet{sdasgupta16} for a critical discussion of how such symmetry demands might be justified. However, this paper is not intended as an exploration of this issue, and I will take the overall cogency of this interpretive constraint for granted in what follows. Were this interpretive principle to be jettisoned, it would have consequences for the physical content of special relativity that go far beyond the relationship between energy and mass.}

Failure to keep track of which individual quantities represent fundamental properties and which do not---of which individual quantities are Lorentz-invariant and which are not---is the core of Lange's argument against mass--energy equivalence. For the mass of a particle is typically represented by the Lorentz-invariant quantity $m$ and is often interpreted as a fundamental physical property, whereas its kinetic energy $(\gamma_v-1)mc^2$ and total free energy $\gamma_vmc^2$ are manifestly \emph{not} Lorentz-invariant (depending as they both do on frame-dependent speed) and thus not candidates for representing fundamental physical properties.\fn{Of course something similar is also true in classical dynamics, as speed isn't Galilean invariant either, but in that context no one alleges an equivalence between mass and energy.} How, then, are we to understand the alleged physical conversion between mass and kinetic energy---a central motivation for the received view---if one quantity in that conversion is fundamental and the other is not? Indeed, in what sense could mass and kinetic energy be ``two forms of the same thing'' if mass is a primitive physical property but energy is not? On the face of it, the widely-publicized equivalence between mass and energy appears to be the product of a ubiquitous conceptual confusion. They simply cannot be equivalent (or interconvertible) because they don't share the same ontological category.\fn{I interpret Lange's characterization of mass as a ``real property'' to be that it is a \emph{fundamental} property. See \citet[p.227]{mlange01}. When indexed to a frame, energy is a perfectly \emph{real} physical property (just like speed). The issue is that it's not fundamental.}

One could simply insist that the quantity $mc^2$ represents a distinct and fundamental type of energy, but it's unclear what the motivations for such a view would be or why the resulting property would warrant the label `energy'. A defining feature of energy is its ability to be converted from one form to another.\fn{See \citet[p.74]{wrindler91}.} What sense does it make to call $mc^2$ `rest energy' if it, but not any other forms of energy, are fundamental physical properties---if we can't make physical sense of its conversion into other forms of energy? Alternatively, one might jettison the idea that mass is a fundamental physical property. Lange's puzzle dissolves if both mass and energy are derivative features of particles. However, in that case one is left wondering exactly what the underlying particle ontology really is. If not in virtue of differing masses, what fundamentally accounts for the different dynamics (e.g., accelerations) of particles subject to identical impressed forces?\fn{Again, it's worth emphasizing that Lange's puzzle raises no concerns about the consistency of $E=\gamma mc^2$ (or $E_0=mc^2$) within the mathematical formalism or about its use in empirically successful applications. Rather, the issue here is a conceptual one about how the theory's ontology is to be understood.}

\section{Descriptive Conversion?}

Lange's solution is to deny the identification of mass as a form of energy on the grounds that there are no genuine physical conversions between mass and energy. Mass is no more a type of energy than photon frequency. The impression of conversion that various physical processes elicit is, on this view, an artifact of our choice of description---a feature of the perspective that we adopt. In this sense, \emph{we} convert energy to mass (or mass to energy), not nature.

The diagnosis here hinges on the observation that mass in special relativity is \emph{non-additive}: the mass of a system is generally not the sum of the masses of its parts. Consider a system of eight particles of equal mass $m$, each moving radially outward with the same speed $u$ from a common origin. This is depicted in the following diagram:

  \begin{center}
	\includegraphics[width=2.5in]{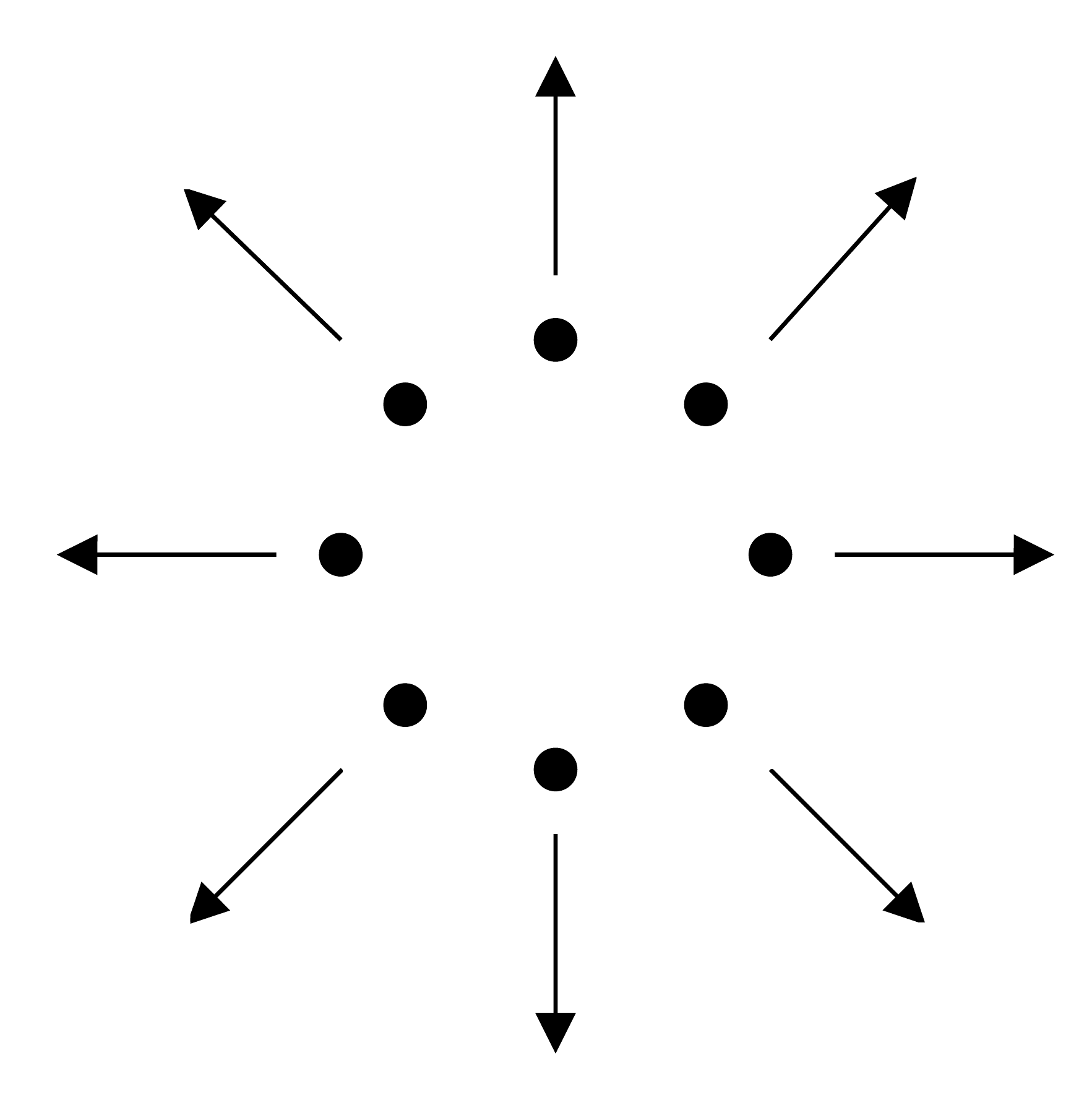}
  \end{center}

\noindent The frame represented in the diagram is the system's rest frame. The system's total free energy in this frame, which is the sum of all eight constituent particles' energies, can thus be expressed as $$E_{sys} = Mc^2 = E_1 + \dots + E_8.$$ But for each particle $i$, $E_i = \gamma_umc^2$, and so we can write $$Mc^2 = 8\gamma_umc^2,$$ from which we conclude: $$M =8\gamma_um > 8m.$$ Here of course there are no collisions or any other postulated interactions, and thus no pretense of converting energy into mass. Instead, the lesson is that the mass of the system is \emph{more} than the sum of its constituent masses. Re-expressing the system mass as\fn{See footnote \ref{kinetic}.} $$M = (m_1 + \dots + m_8) + \frac{1}{c^2}(T_1 + \dots + T_8)$$ makes explicit how the system mass depends upon the kinetic energies of its parts. Notice that if we were to boost each particle's speed in equal measure, $u \rightarrow u^\prime > u$, the mass of the system would increase: $$E^\prime_{sys} = M^\prime c^2 = 8\gamma_{u^\prime}mc^2$$ $$M^\prime = 8\gamma_{u^\prime}m > 8\gamma_u m$$ because $\gamma_{u^\prime} > \gamma_u$, and yet the fundamental particle masses would remain fixed.

This example might reasonably make us suspicious regarding claims of mass--energy interconversion. The special relativistic formalism tells us that the system mass goes up when the particles are uniformly boosted in their respective directions, and thus it appears energy is being converted into mass. Yet when we look at the particles themselves only their kinetic energies are changing. Their masses remain fixed, and so it looks like whatever energy is applied to the system in the boost is converted into the kinetic energies of the constituents. At no point does the mass of any electron in this process change, for example, so in what sense is mass being gained or lost?

To make sense of these and other apparent cases of mass--energy conversion, Lange distinguishes two ways we might characterize a system of particles. First, there is the `component level' description. Here the collection is described in terms of its constituent particles and the forces and interactions they experience. Second, there is the `system level' description, where we think of the collection as a single unit interacting with an external environment and ignore the internal dynamics. We shouldn't be misled into thinking each description is on an ontological par: the component level is fundamental. While it is true that energy is apparently converted into mass when a ball of gas is heated (to use Lange's example), this `conversion' only occurs when the system-level description is invoked.\fn{\citet{bondispurgin87} use this example to argue that energy \emph{has} mass. Read in a straightforward way, this claim is confused. Mass is a property of inertial resistance: the mass of an object is a measure of how much that object resists changes to inertial motion in light of impressed forces. So for energy to \emph{have} mass, energy must be the sort of thing to which impressed forces can be applied, and energy simply isn't that sort of thing.} Described entirely in terms of basic constituents, Lange argues, there is no conversion between mass and energy---just as there is no conversion in our initial example when all eight particles are boosted. This is why he writes that mass--energy conversion is an artifact of our perspective and not a real physical process.

Other cases involving particle systems mislead us into thinking that energy--mass conversion occurs because we inadvertently switch perspectives midway through our analysis. For instance, consider how Lange diagnoses the typical textbook treatment of an inelastic collision. Recall from section \ref{inelastic} that the kinetic energies of the pre-collision bodies appeared to be converted into mass upon impact, as outlined in the diagram below:

  \begin{center}
	\includegraphics[width=3in]{Diagram1}
  \end{center}
  
  \vspace{0ex}
  
  \begin{center}
  	\includegraphics[width=3in]{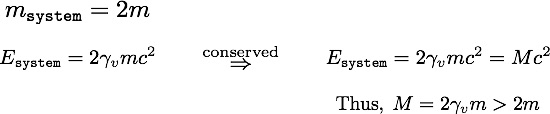}
  \end{center}

\noindent But the total pre-collision mass is $2m$ only if we adopt a component-level description. As one can see in the following diagram, on a system-level description (owing to non-additivity) the pre-collision mass is $2\gamma_vm$---precisely the mass of the system after collision:

  \begin{center}
	\includegraphics[width=3in]{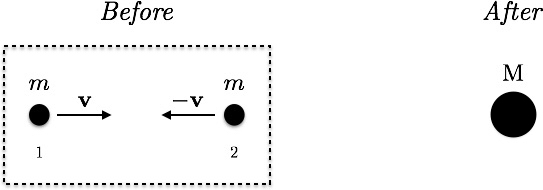}
  \end{center}
  
  \vspace{0ex}
  
  \begin{center}
  	\includegraphics[width=3in]{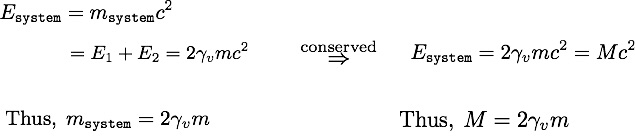}
  \end{center}

\noindent The \emph{appearance} of conversion from energy to mass in this instance arises because textbooks switch descriptive levels halfway through the analysis. If one describes the process from a fixed perspective, there is no conversion. A similar diagnosis applies to the apparent mass defect involved in tritium decay, although there the descriptive switch occurs in the opposite direction: mass appears to be converted into energy because our pre-decay description is at the system-level, whereas our post-decay description is at the component level.\fn{One worry with Lange's account, raised by \citet{fflores05}, is that the underlying ontological picture arises from inconsistent application of the relevant interpretive principles. Lange's use of `mass' to designate an object's \emph{rest} mass \citep[p.225]{mlange01} appears to implicitly privilege a particular frame---namely, the frame in which the object is (instantaneously) at rest. Like length, then, it is a perfectly real physical property, but on the surface ought to be no more fundamental than particle mass in any other inertial frame---which is to say, not fundamental. However, it is clear from Lange's discussion of rest mass and relativistic mass \citep[pp.226--7]{mlange01} that rest mass is understood merely as the $u = 0$ mathematical limit of relativistic mass. As previously noted (see note \ref{flores1} above) the quantity $m$ itself represents an ontologically fundamental and frame-invariant particle property for Lange, which also happens to be equal to the particle's relativistic mass in the frame in which the particle is at rest. But that equality is not constitutive of the property. In this way, contra \citet{fflores05}, Lange's proposed ontology preserves the special relativistic maxim that all inertial frames are physically equivalent. \label{flores2}}

\section{Puzzles of Dynamical Interpretation}

However, Lange's diagnosis of alleged mass--energy conversions fails to convince. In this section I explain why. In the process I identify several puzzles we must confront in light of adhering to Lorentz invariance as an interpretive constraint in understanding the fundamental ontology of special relativistic particle dynamics. These puzzles then help to motivate the account of what makes special relativity dynamically novel that I develop in the final sections---an account that makes no mention of any equivalence between mass and energy.

\subsection{Elementary Conversions}

Assume for the moment that Lange has correctly diagnosed the alleged cases of mass--energy conversion discussed above and that such conversions are, in fact, unphysical. An initial problem is that his analysis doesn't generalize. There are a wide class of particle collisions where apparent energy--mass conversion can't be explained away as a perspectival shift. The most obvious instances are electron--positron creation and annihilation: $$\gamma + \gamma \rightarrow e^- + e^+ \textrm{\qquad (electron--positron creation)}$$ $$e^- + e^+ \rightarrow \gamma + \gamma \textrm{\qquad (electron--positron annihilation)}$$ In the first case, two high-energy massless photons collide to create an electron and a positron, both elementary particles of equal non-zero mass. In the second, the reverse happens. In both instances there is an apparent conversion between mass and energy that can't be explained away by appealing to different perspectives or levels of description. The pre-collision situation is characterized in terms of two distinct things, as is the post-collision situation. But in the creation reaction the input bodies \emph{have no mass}, whereas the output bodies do. Where has the mass come from? Certainly not from switching our perspectives part way through the analysis. Some of the photons' energy really has, it would seem, been converted into mass. Indeed, the kinetic energy difference pre- and post-collision precisely matches the energy associated, via $\Delta E_0 = \Delta mc^2$, with the electron and positron masses. Contra Lange (especially \citet[p.230]{mlange01}), there is no plausible case to be made here that mass alone is conserved. It is total energy that's conserved in this reaction, where that total energy would seem to include mass as one particular form.

These examples are not isolated or unique. A host of particle reactions can't be accommodated within Lange's analysis: $$\pi^0 \rightarrow \gamma + \gamma \textrm{\qquad (neutral pion decay)}$$ $$\gamma + p \rightarrow \pi^0 + p \textrm{\qquad (pion photoproduction)}$$ $$p + p \rightarrow p + p + p + \bar{p} \textrm{\qquad (proton--antiproton pair production)}\fn{This terminology follows \citet[p.247]{jfreund08}.}$$ $$e^- + e^+ \rightarrow e^- + e^+ + e^- + e^+ \textrm{\qquad (electron--positron pair production)}$$ In each, the apparent conversion between energy and mass occurs at the fundamental descriptive level, so no plausible story about shifting perspectives is available. These reactions really do seem to involve mass coming into and out of existence.

Unfortunately, the existence of these reactions does little to quell the original sense of puzzlement. In the particle reactions it seems hard to deny that mass is created and destroyed, and mass--energy conversion in accordance with $\Delta E_0=\Delta mc^2$ is the natural explanation. But how is such a story conceptually coherent if mass is a fundamental physical property and energy (being Lorentz-\emph{variant}) is not?

Here we must be careful to manage expectations and to distinguish two issues. There is, on the one hand, the physical account of what's going on in these reactions and why they occur. What physical laws govern electron-positron annihilation and creation, for example? We should not expect special relativity (or at least special relativistic particle dynamics) to answer this question. The collision reactions at issue are inherently quantum in nature and our best understanding of them lies in quantum field theory, not in the non-quantum particle dynamics of special relativity.

On the other hand, quantum field theory is designed to preserve central conceptual and dynamical features of special relativity, such as the physical equivalence of all Lorentz frames, and the collision reactions noted in this section \emph{are} often cited in support of the purely special relativistic identification of mass with energy.\fn{Many of the core classical principles of relativistic particle dynamics are strikingly well-confirmed by the very reactions at issue. See \citet{afrench68} for a discussion of some of the experimental evidence from particle physics for the basic principles of special relativistic particle dynamics.} Is there a way of understanding the fundamental ontology of special relativistic particle dynamics that remains \emph{consistent with}---even if not \emph{explanatory of}---the existence of these collision reactions, without thereby either committing the conceptual error of identifying fundamental and derivative physical properties or giving up the principle that all inertial frames are physically equivalent? That seems like an entirely reasonable question. What Lange's puzzle makes vivid is that standard textbook accounts, according to which there is simply a conversion between mass and energy, fail to provide this.

\subsection{Composite Mass}

Let us back up a step: is Lange's analysis adequate for the cases he considers explicitly? Return to his diagnosis of the ball of gas and recall that, when the gas is heated, there is an apparent conversion of energy into mass. Lange's claim is that this conversion is unphysical and the product of a particular perspectival switch, namely, from that of treating the ball of gas as composed of distinct molecules to that of treating the ball of gas as a single unit. When the gas is viewed from the more ontologically fundamental perspective---that of the gas molecules themselves---the application of heat merely changes the kinetic energies and not the molecular masses. On this basis, Lange concludes that ``this `conversion' of energy into mass is not any kind of real physical process taking place in nature. \emph{We} `converted' energy into mass simply by \emph{changing our perspectives} on the gas: from treating it as many bodies to treating it as a single body'' \citep[p.235]{mlange01}.

This diagnosis only makes the situation of the gas more mysterious, however, for there \emph{is} a genuine and fundamental physical process occurring here in need of identification. If we imagine the gas (as a system) to be uniformly distributed in an enclosing box or shell, that box will be harder to move---will exhibit more inertial resistance to impressed forces---after it's been heated. This is a real empirical effect, not a magician's conjuring trick!\fn{See \citet{bondispurgin87}. This point is also noted in \citet{fflores05}. A similar phenomena occurs in a compressed spring, which has a larger mass than a relaxed spring owing (it would seem) to its increased potential energy. On this point, see \citet{cdib13}.} There is an undeniable physical change in the gas' total mass after heating, and that change must be grounded in changes to the fundamental properties of the basic entities constituting the gas. That is, there is a genuine physical fact about the inertial resistance the gas puts up---a fact that is grounded in the fundamental properties of the gas' basic constituents---and thus a genuine physical fact about whether that mass is changing. It is! That the gas' mass (as a system) can change while the masses of its molecular constituents remain constant is in fact precisely what non-additivity shows.

Therein lies the real mystery, the one that Lange's analysis obscures. A physical change in the gas' mass is clearly occurring and that change must be grounded in changes to the fundamental properties of the gas' elementary constituents. But how can that be if kinetic energy isn't a fundamental physical property and if the kinetic energies of the constituent molecules are the salient things changing? If energy is not being converted into mass, what exactly is happening? (Put another way, how do we make sense of this process as a genuinely physical one in light of non-additivity?)\fn{This puzzle---the puzzle of understanding the nature of composite mass---is not unique to macroscopic objects or objects whose constituents only interact via collisions (as one generally assumes for gases). For all but the most fundamental particles, mass seems to be partly constituted by `internal' energy, whether in the form of kinetic energy or some form of binding or potential energy.}

Turn now to Lange's analysis of inelastic collisions. The charge was that in the process of describing the collision we inadvertently switched perspectives, and in the process `converted' the kinetic energy of the initial bodies into the rest mass of the final body. If we think of the two pre-collision bodies as a single system, the total system mass will equal the mass of the single body post-collision and the alleged conversion evaporates. Again (the claim is) the conversion is revealed to be unphysical, the result of descriptive legerdemain.

But for this analysis to be compelling, Lange must do more than show that the mass of the pre-collision two-body arrangement, when viewed as a single system, is the same as the mass of the post-collision body. He must (in the first instance) establish that such descriptions amount to characterizing the pre- and post-collision arrangements \emph{from the same perspective}, and that requires saying a good deal more about how exactly these `perspectival descriptions' are to be understood. What justifies Lange's claim that the textbook analysis \emph{switches} perspectives? That we view the pre-collision arrangement as two bodies and the post-collision arrangement as one? Am I switching perspectives when I smash a rock and describe the result as a collection of shards?

More importantly, though, Lange's diagnosis fails in this case for the same reason it did for the ball of gas. There is a genuine physical process occurring here in which mass is changing---a process that is grounded in changes to the fundamental properties of the particles that constitute the colliding bodies. Imagine that two equally massive balls of putty collide in an inelastic collision. If I then cut the post-collision body in half, each will have a greater mass than either pre-collision body. This is a real physical change and Lange's talk of `switching perspectives' obscures it. The obvious explanation of this change is that the kinetic energies of the putty molecules have increased, but now we're back at non-additivity and the ball of gas puzzle.

\section{An Ontology for Relativistic Particle Dynamics}

The situation we now face is the following: there are compelling grounds to deny the inter-convertibility and equivalence of mass and energy, and yet such a denial seems hard to reconcile with important experimental phenomena. What should we make of this? I suggest that the difficulties stem from an inadequate (or inadequately specified) ontological picture of special relativity and from a hitherto overlooked feature of its dynamical foundations. Accordingly, this section develops a new account of the fundamental ontology of special relativistic particle dynamics. This requires identifying the metaphysically basic entities and properties, indicating how they interact and evolve dynamically, and specifying the mathematical objects and structures and equations that encode that physical picture.\fn{The approach here is in the spirit of \citet{tmaudlin18}.} The ontology is based on a generally covariant (or geometrical) formulation of special relativity, briefly outlined in the first subsection. Several independent motivations for my ontological picture are offered, although the central claim---developed in sections 7 through 9---is that this interpretation provides a satisfying resolution of the puzzles associated with mass--energy `equivalence' while clarifying the dynamical foundations of special relativity and the meaning of $E_0=mc^2$.

It is now commonplace to formulate special relativity in 4-dimensional terms, so one might not think this section has anything new to offer.\fn{Standard philosophical references include \citet{mfriedman83}, \citet{jearman89}, and \citet{dmalament12}.} But: (1) this formulation is rarely accompanied by any discussion of the fundamental ontological picture associated with the dynamics; (2) there has been no discussion in the literature of how the experimental phenomena associated with alleged energy-mass conversion ought to be understood in geometrical terms; and, most importantly for foundational concerns, (3) despite the ubiquity of 4-dimensional presentations, it has yet to be recognized---as I think it ought to be---that the central \emph{dynamical} insight of special relativity has nothing to do with the relationship between energy and mass. Indeed, 4-dimensional expositions of special relativity remain frustratingly elusive regarding what the central dynamical (as opposed to kinematical) insights really are.

\subsection{The Dynamical Formalism on Minkowski Spacetime}

Geometrically, special relativistic theories are framed against the backdrop of Minkowski spacetime, which is a space of events represented by a 4-dimensional (pseudo-Riemannian) manifold equipped with a flat metric $\eta_{\mu\nu}$ of (Lorentzian) signature $(-,+,+,+)$.\fn{Where there is little risk of confusion I will be lax about the distinction between mathematical representation and physical feature represented.} The trajectory or worldline of a material particle is the collection of events that together constitute the history of that body, and it is a basic postulate of special relativity that such objects are represented by timelike worldlines and light rays by null trajectories.\fn{Recall that the metric structure divides the spacetime at any point (call it the `origin point') into distinct regions, which can be characterized by the vectors (4-vectors) at that point. The \emph{timelike} region consists of those events whose displacement vectors from the origin point have negative magnitude. All such events are said to be timelike separated from the origin point and any 4-vector that points from the origin point to a timelike separated event is said to be a timelike 4-vector. The \emph{lightlike} (\emph{spacelike}) region is the set of events whose displacement 4-vectors from the origin point are of null (positive) magnitude. This definition extends to 4-vectors as above. A curve through the manifold is said to be timelike (null, spacelike) if the tangent 4-vector at each point along it is timelike (null, spacelike).

Note that photons play a rather curious role in textbook presentations of special relativity, with some authors smoothly sliding between initial talk of light rays or signals to later talk of photons and other authors acknowledging that photons are quantum mechanical in nature and thus not a part of special relativistic dynamics proper. (Compare \citet[pp.84--86]{wrindler91} and \citet[p.173]{vfaraoni13}.) The latter view is how special relativity was originally understood: in the 1920s, well after the acceptance of special relativity, Bohr and others continued to express doubts about the existence of photons. See, e.g., \citet[pp.230ff]{apais91} and \citet[pp.19ff]{dmurdoch90}. The presentation here is deliberately silent on the behavior of photons in Minkowski spacetime. Indeed, as \citet[p.8]{wrindler91} notes, light itself is not essential to the spacetime structure of special relativity. \citet[Ch.4]{tmaudlin12} makes a related claim, developing the kinematics in a way that makes no mention either of inertial frames or of light's constitution and `speed'.
}

Each point along a particle's worldline is associated with a unique 4-vector $P^\mu$ called the \emph{4-momentum} (or \emph{energy-momentum 4-vector}), which is always tangent to the curve and is traditionally defined in textbooks as $$P^\mu = mU^\mu,$$ where $$U^\mu = \frac{dx^\mu}{d\tau}$$ is the 4-velocity along the worldline.\fn{Some authors use bolded capital letters (e.g., $\vec{A}$) for 4-vectors and bolded lower-case letters (e.g., $\vec{a}$) for spatial 3-vectors, indicating the components in an inertial frame $S$ by writing $\vec{A} \stackrel{S}{\rightarrow} (A^0,A^1,A^2,A^3)$ or $\vec{a} \stackrel{S}{\rightarrow} (a_1,a_2,a_3)$. I will occasionally use this notation, but more often will refer to 4-vectors by writing things like $A^\mu$, understood to represent the components of the 4-vector $\vec{A}$ in some arbitrary inertial frame.} $\tau$ is the proper time parameter along the particle's worldline. In an arbitrary inertial frame the 4-momentum can be expressed in coordinate form as: $$P^\mu = (\gamma_u mc, \gamma_u m\vec{u}) = (E/c,\vec{p}),$$ where $\vec{u}=\frac{d\vec{x}}{dt}$ is the spatial velocity in that frame, $E=\gamma_umc^2$ is the relativistic energy of the particle, and $\vec{p}=\gamma_um\vec{u}$ is its relativistic spatial momentum (or relativistic 3-momentum). These features are illustrated in the following spacetime diagram: 
\begin{center}
	\includegraphics[width=3in]{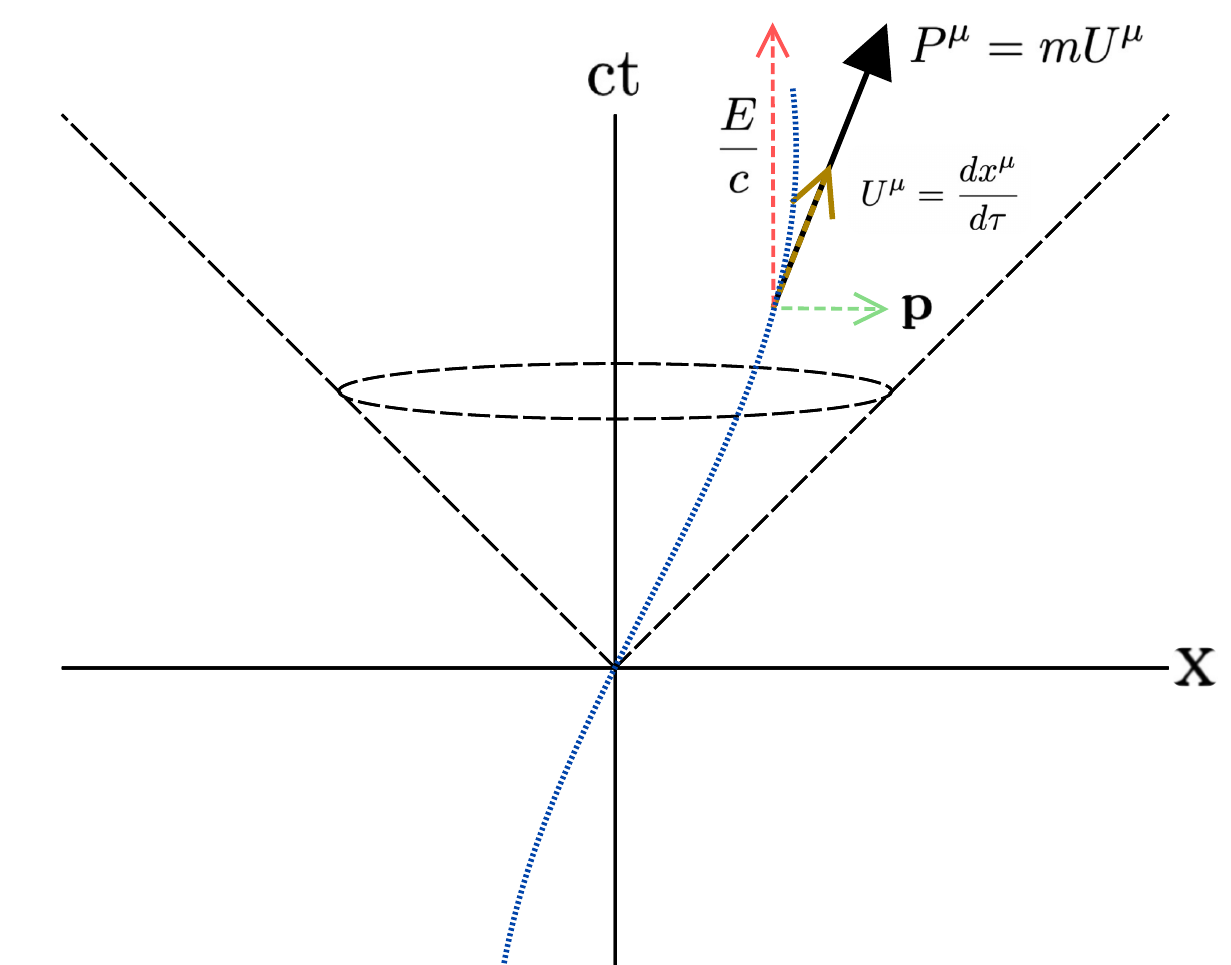}
\end{center}
Here it's worth emphasizing that both relativistic energy and mass show up as components in the frame-dependent decomposition of the 4-momentum, so we might expect there to be some connection between 4-momentum and a clear understanding of the relationship between energy and mass.

There are several dynamical principles governing special relativistic particles. First, in response to a 4-force $F^\mu$ a particle obeys what appears to be a 4-dimensional relativistic analogue of Newton's equation of motion: $$F^\mu = \frac{dP^\mu}{d\tau}.$$ The components of the 4-force in an arbitrary inertial frame are given by $$F^\mu = \gamma_u (c \frac{d}{dt}[\gamma_u m], \frac{d\vec{p}}{dt}) = \gamma_u (\frac{1}{c} \frac{dE}{dt}, \vec{f}).$$ Here $\textbf{f}$ is the relativistic 3-force $$\vec{f} = \frac{d\vec{p}}{dt} = \frac{d(\gamma_u m\vec{u})}{dt},$$ which in the $u/c \rightarrow 0$ limit reduces to what one might classically identify as the Newtonian force. 
Second, the total 4-momentum of a set of colliding particles, $\vec{P}_{\mathrm{tot}}$, is always conserved: $$\vec{P}_{\textrm{tot}} = \sum_{i=1}^n \vec{P}_{(i)} = \textrm{constant},$$ where the $\vec{P}_{(i)}$ are the 4-momenta of the incoming (or outgoing) particles. This is an independent dynamical principle and not something derived from a more basic law. Let us call it the \emph{4-Momentum Collision Principle}.\fn{\label{nofields}See \citet[pp.70--73, 90--92]{wrindler91}. For an isolated $n$-body system that interacts only locally (i.e., effectively via collisions), a more general principle holds that the net 4-momentum \emph{of the system} remains constant, in the sense that the components of the total system 4-momentum do not change in any given inertial frame. Even though the specific $\vec{P}_{(i)}$ 4-vectors that are elements of this sum are frame-dependent because the summation is taken at an instant, in these circumstances \citet[pp.78--79]{wrindler91} shows that the total system 4-momentum $\vec{P}_\textrm{sys}$ is a well-defined 4-vector. It follows from this more general principle that relativistic energy and relativistic (spatial) momentum are both independently conserved in these circumstances, even though the values of those quantities are frame-dependent. For a more general discussion not restricted to inertial observers, see \citet[pp.288--291]{egourgoulhon10}. \emph{When an $n$-body system can't be treated as isolated, such as when various sorts of field-theoretic considerations are included, then there isn't generally a well-defined total 4-momentum associated with the system.} However, the puzzles developed above concern physical systems for which these field-theoretic considerations can be ignored. Recall (see footnote \ref{method}) that the guiding methodology here is to tease out the significance of $E_0=mc^2$ in the simplest dynamical cases and to leave more complicated physical situations for subsequent work.} Together, I will argue, these two dynamical principles---when coupled with the right interpretation of the theory's fundamental ontology---clarify the physical significance of $E_0=mc^2$ and suggest an alternative picture of what makes special relativity dynamically novel.\fn{There is a third dynamical principle---conservation of the angular momentum 4-tensor along any particle's world line---but it won't play a role in the argument that follows.}

\subsection{Outline of a New Ontology}

What metaphysics ought to be associated with this formalism? Standard presentations of \emph{classical} dynamics encourage a particular ontological picture, according to which there are (a) material particles possessing primitive properties of mass and instantaneous location and velocity in space or spacetime, and (b) forces that mediate interactions between particles and generate changes in their dynamical states and motions.\fn{Roughly speaking, a \emph{dynamical state} of a particle is the collection of fundamental properties relevant to the types of interactions it generates and the way in which it responds to different types of interactions. A particle may possess properties that are relevant to \emph{whether} it experiences a particular impressed force or interaction, but which are not part of its dynamical state---e.g., its position in space or spacetime. A non-interacting object moving inertially is constantly changing its position and yet its dynamical state remains constant.} Other quantities of physical significance---e.g., ``dynamical variables'' like spatial momentum, angular momentum, and energy---are ultimately understood as derivative properties grounded in, or defined in terms of, what's ontologically fundamental.\fn{See, e.g., \citet[pp.13--14]{josesaletan98}. I do not mean to suggest that this picture of classical ontology is uncontroversial. Among the issues raised in recent years, some philosophers have considered whether velocity ought to be taken as an ontologically primitive property in its own right (see \citet{farntzenius00}, \citet{jcarroll02}, \citet{umeyer03}, \citet{ssmith03}, \citet{mlange05}, \citet{keaswaran14}, \citet{cmccoy18}), whereas others have wondered whether attributions of mass ought to designate fundamental properties of particles (see \citet{sdasgupta13}, \citet{nmartens19a}, \citet{nmartens19b}). And \citet{jbutterfield06} has argued against understanding classical dynamics (or indeed any physical theory) in terms of properties defined at points of space or spacetime. There is also a long tradition within the metaphysics of physics going back at least to Mach and Hertz puzzling over just what sort of thing a \emph{force} really is. For more recent discussion, see, e.g., \citet{bellis76}, \citet{bigelowellispargetter88}, and \citet{jwilson07}. Some of the issues raised in these literatures will be relevant to the specific way I frame the ontological proposal sketched below, but they do not affect the central argument and in the context of this paper I've had to set them aside.}

On the other hand, while typical presentations of special relativity often contain extensive discussions of the radical ontological changes brought about by relativistic \emph{kinematics} (e.g., the relativity of simultaneity, length contraction, time dilation), they remain surprisingly quiet regarding any ontological changes demanded by the new \emph{dynamics}---aside, as we've seen, from problematic claims about the equivalence of mass and energy. Indeed, the very manner in which the new dynamics is often introduced as ``a modification of the Newtonian scheme'' and the similarity in terminology gives the impression that much of the ontology standardly associated with classical dynamics carries over, more or less, to the context of special relativistic particle dynamics.\fn{\citet[p.4]{afrench68}} We are told, for example, that classical momentum ($\vec{p}_{\textrm{cl}} = m\vec{u}$) must be `redefined' in the relativistic context and that the equation $\vec{f} = \frac{d\vec{p}}{dt}$ is the `relativistic generalization' of Newton's second law. Many relativistic expressions appear as seemingly straightforward analogues of Newtonian ones, thereby suggesting that the basic ontology underpinning the dynamics remains essentially the same.\fn{See, e.g., \citet[pp.21--23]{afrench68}.}

This impression strikes me as profoundly mistaken. Relativistic particle dynamics requires a radically different fundamental ontology from the one usually associated with classical dynamics---one that goes well beyond differences associated with kinematics. I think the correct ontology is based on the 4-dimensional covariant formulation of the dynamics, and that the failure to realize this (or at least to articulate clearly that ontology) obscures the solution to Lange's puzzle and the novel foundational features of the dynamics itself.

\subsubsection{Dynamical States of Particles}

If we take the geometrical formulation of special relativity seriously and understand the dynamics as genuinely unfolding in Minkowski spacetime, the 4-momentum vector is the obvious candidate for encoding a particle's instantaneous dynamical state. There is a natural ontology associated with this, according to which the 4-momentum encodes two distinct and fundamental properties of a particle: one represented by the 4-momentum's \emph{magnitude} and the other represented by what I will call the 4-momentum's \emph{orientation}. (This is not to deny that 4-momentum components represent objectively real particle properties when indexed to a particular frame, but those features are not fundamental.) For material bodies, whose worldlines are timelike, these properties are physically independent. So the proposal is that, at any given location in spacetime at which a massive particle exists, this pair collectively defines the instantaneous dynamical state of that particle at that location.

As ontologically primitive features of particles, these properties are not subject to reductive analysis or further explication, although one can still convey a feel for their physical content. The magnitude of a particle's 4-momentum is in some sense a reflection of how much the particle resists changes to its dynamical state on account of an applied 4-force. This quantity is often taken to be just another way of representing particle mass, an impression reinforced by the fact that $\| P^\mu \| = \sqrt{-P_\mu P^\mu} = mc$ holds as a fixed relation for all material particles. I have no objection to this conceptual gloss on the physical content of 4-momentum magnitude, but in keeping with my emphasis on the covariant formalism I prefer to think of $P^\mu$ as the central representational device: the customary definition of 4-momentum, $P^\mu = mU^\mu$, is, on this view, a definition of the quantity $m$.\fn{There is some precedent in the physics literature for thinking of $m$ this way. See, e.g., \citet[p.272]{egourgoulhon10}. I do not think anything of deep philosophical substance hangs on this point: both $\| P^\mu \|$ and $m$ are understood as representing one and the same primitive physical property. However, within the special relativistic formalism the quantity $m$ is most salient in frame-dependent contexts, where it plays a central role in dynamical equations governing \emph{derivative} ontology. Since I think the focus on the frame-dependent equations for special relativistic particle dynamics has been the source of much ontological confusion, I prefer to do without $m$ and use $\| P^\mu \|$ as the representation of particle mass.

It's worth emphasizing that the property represented by $m$ (or $\| P^\mu \|$) takes on a rather different character in relativistic dynamics than it does in classical dynamics. Within an inertial frame a special relativistic particle, unlike a classical one, exhibits different amounts of inertial resistance to impressed 3-forces in different directions. The resistance to being accelerated in a direction parallel to a particle's instantaneous spatial velocity---its `longitudinal mass'---is different from its resistance to being accelerated in a direction perpendicular to its instantaneous spatial velocity---its `transverse mass'. Indeed, the spatial acceleration of a body in response to an impressed 3-force is generally not even in the direction of the 3-force itself. The frame-dependent dynamical equation governing particle motion is $\vec{f} = \gamma_um\vec{a} + \frac{d}{dt}[\gamma_um]\vec{u}$, and so the magnitude and direction of a particle's acceleration in response to an impressed force depends on properties other than just $m$ and the direction in which the force is applied (e.g., its spatial velocity in a frame). The relationship $\vec{f}=m\vec{a}$ holds only in the instantaneous rest frame of the particle. \citet[pp.195--198]{jfreund08} describes a very simple example where a constant 3-force applied to a particle solely in the x-direction generates a velocity-dependent deceleration in the y-direction.} $P^\mu$ directly represents two basic dynamical properties of particles and is not to be understood as representing derivative features characterized by constellations of mathematical objects that themselves represent more fundamental physical properties.

The second primitive property, 4-momentum \emph{orientation}, reflects roughly how a particle is `directed in' spacetime.\fn{Like $\| P^\mu \|$ and $m$, $P^\mu$ and $U^\mu$ are equally good representations of a particle's (4-momentum) orientation.} Together, we can think of 4-momentum magnitude and orientation as reflecting how a particle is `moving through' spacetime. Although spacetime diagrams may suggest that orientation is an artifact of one's choice of coordinates, it is in fact only our representation that is coordinate-dependent. The orientation itself, \emph{in spacetime}, is entirely independent of any coordinate system.\fn{This sort of discourse naturally suggests an underlying substantivalism of some form or other. I embrace this, but wish to remain agnostic here regarding whether such a metaphysical commitment is necessary for the ontology I propose.} This is of course not the only respect in which spacetime diagrams of Minkowski spacetime can mislead. The following two diagrams are equally adequate graphical representations of the same 4-momentum (same magnitude, same orientation), just drawn from the perspective of different Lorentz frames:
\begin{center}
	\includegraphics[width=3in]{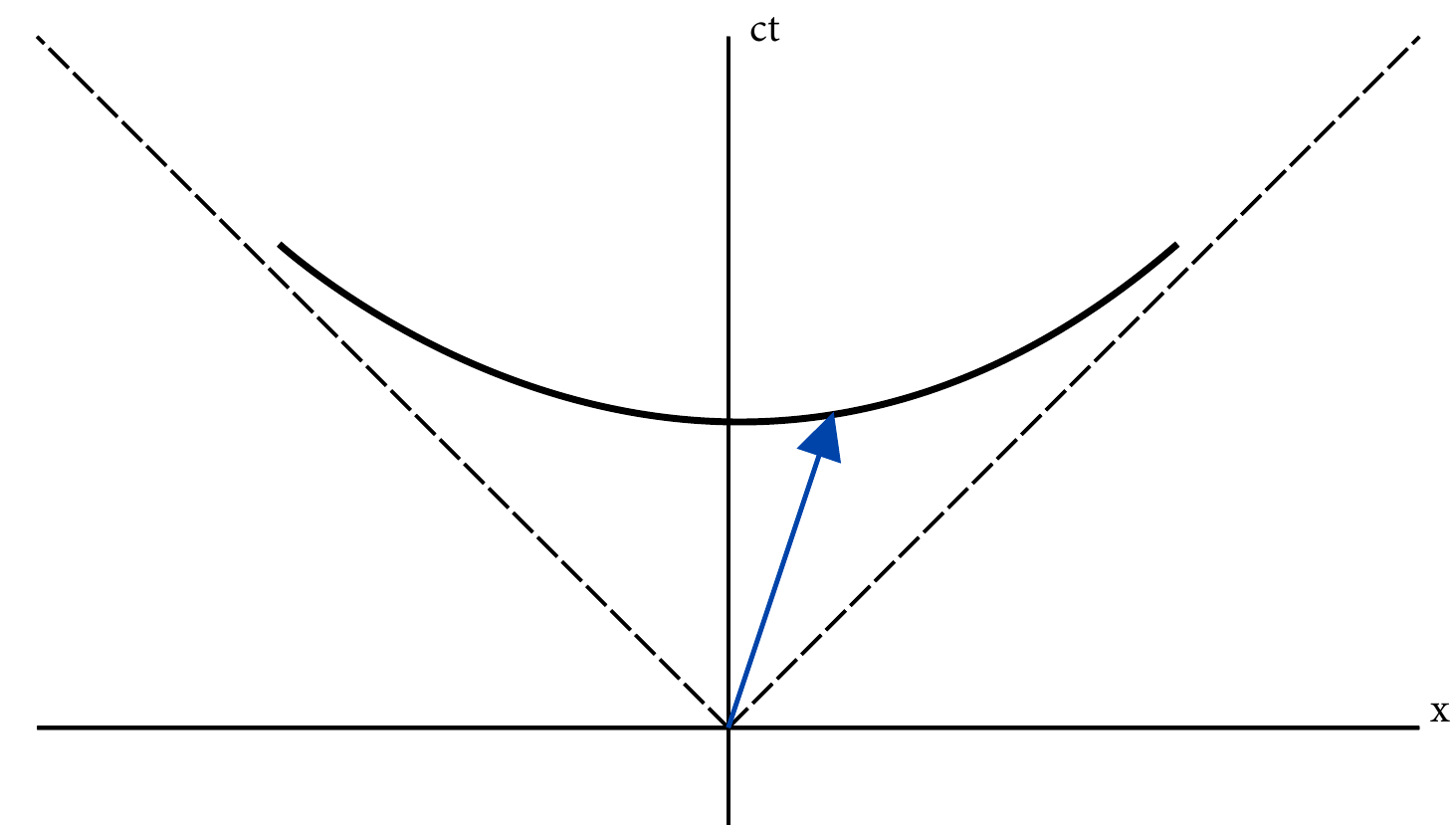}
\end{center}

\noindent and

\begin{center}
	\includegraphics[width=3in]{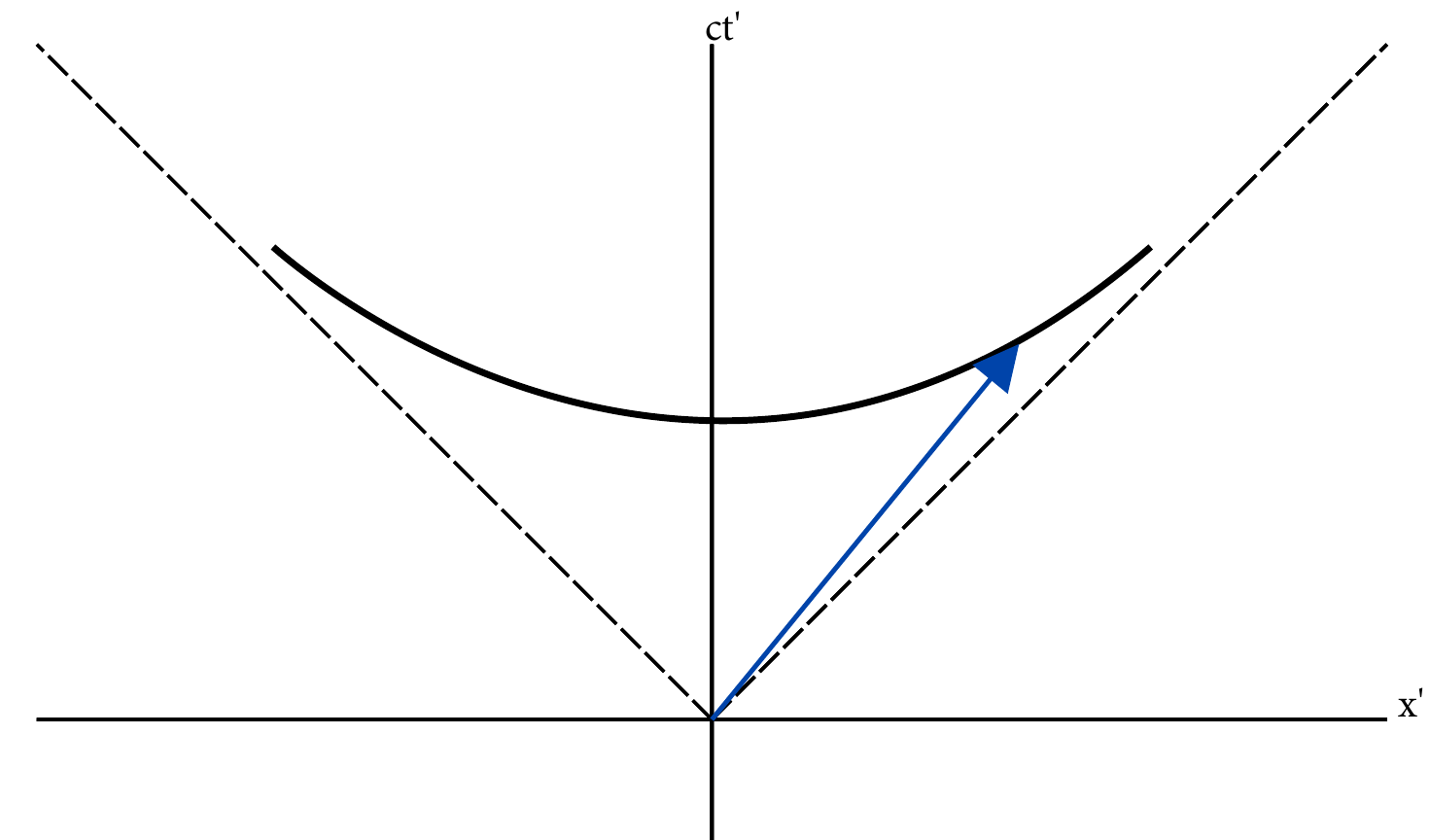}
\end{center}

\noindent On the other hand, the two 4-momenta in the following diagram have the same magnitude but different orientations, as they are being represented with respect to the same Lorentz frame:

\begin{center}
	\includegraphics[width=3in]{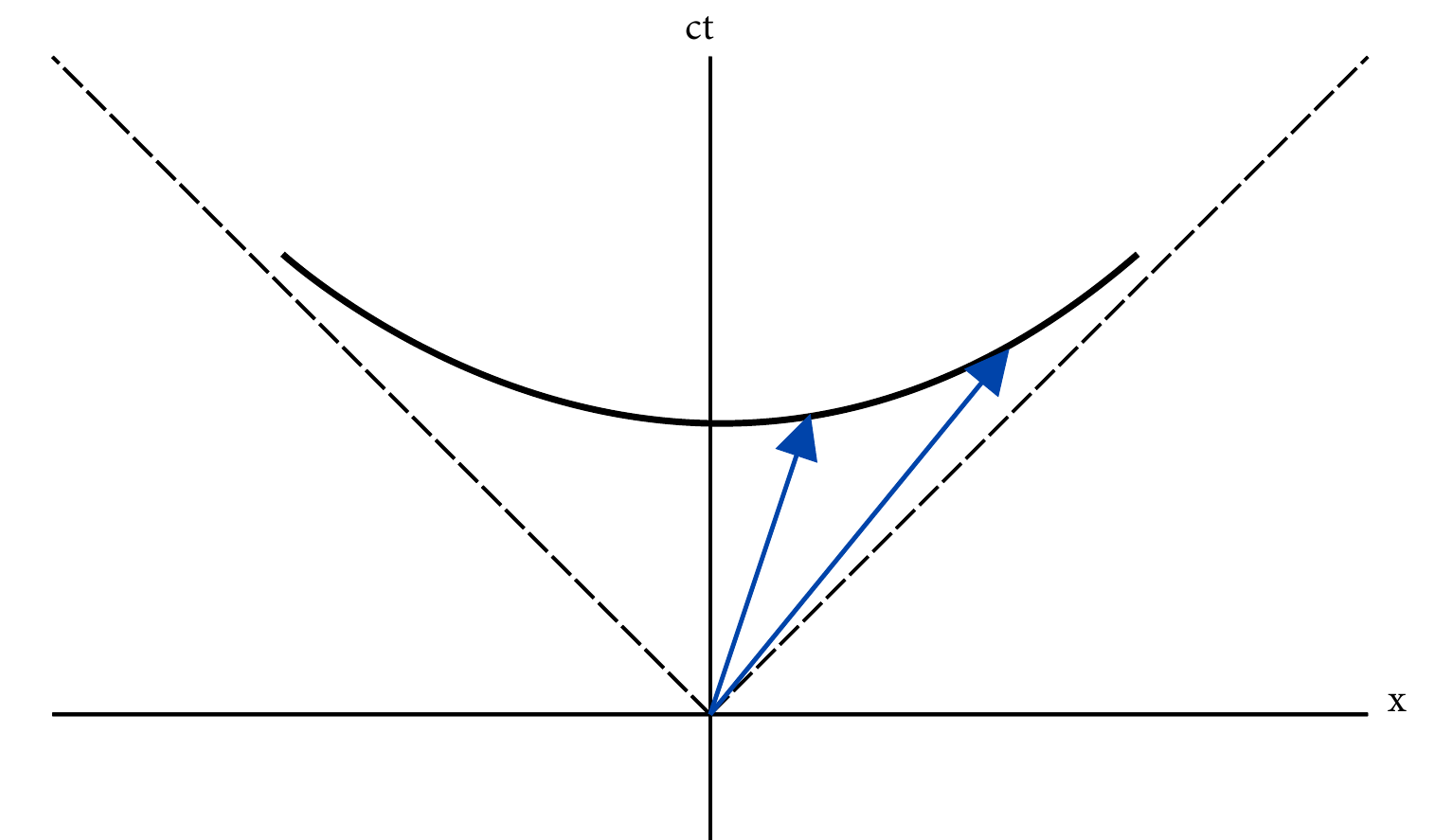}
\end{center}
We cannot specify the 4-momentum orientation of a particle except by invoking indexical expressions---say, by specifying a coordinate system using indexicals and then giving the 4-momentum components in that coordinate system---much as the proponent of substantival space can't specify the absolute location of a particle except by saying (pointing) that it is \emph{here} or \emph{there} or providing its coordinates in some coordinate system.\fn{For more on the role of indexicals in this sort of context, see \citet[pp.189--191]{tmaudlin93}. There are, of course, easily expressible facts about \emph{relative} differences in 4-momentum orientations between particles, as $\vec{P}_1\cdot\vec{P}_2 = \| \vec{P}_1 \| \| \vec{P}_2 \| \gamma(v)$ holds invariantly (where $v$ is the relative velocity between the two particles). See \citet[p.76]{wrindler91}. If $\vec{P}_1\cdot\vec{P}_2 = \| \vec{P}_1 \| \| \vec{P}_2 \|$ then both 4-momenta have the same orientation, and if $\vec{P}_1\cdot\vec{P}_2 \neq \| \vec{P}_1 \| \| \vec{P}_2 \|$ then the extent to which $\vec{P}_1\cdot\vec{P}_2 / \| \vec{P}_1 \| \| \vec{P}_2 \| > 1$ provides a measure of how their orientations differ.} But that alone is not a reason to reject it, especially if its inclusion in the ontology can be given a compelling theoretical justification.

By way of such a theoretical justification, consider a particle moving inertially that is suddenly, for a temporal interval $\Delta t$ as measured in a particular Lorentz frame, subjected to a constant impressed 3-force in the direction of its motion. During this interval the particle accelerates, and then returns to inertial motion after the force is switched off. Intuitively, the dynamical state of the particle has changed: the particle's state before the applied force is switched on is different from its state after the force is switched off. After all, if there were a second (force-free) particle originally at rest with respect to the first, after $\Delta t$ it would be in uniform relative motion with respect to it (or in a different state of relative motion if the second particle wasn't originally at rest). But what about the \emph{dynamical state} of the particle itself has changed? What fundamental physical properties does it have that we can point to as having changed from one side of this interval to the other? All of the obvious candidates---kinetic energy, speed, 3-momentum magnitude---aren't Lorentz-invariant and thus aren't fundamental, and the orientation of its 3-momentum remains unchanged because the force is applied in the direction of motion. The dynamical state of the particle has changed, but when considered within a frame it's not at all clear exactly \emph{which} fundamental properties have changed.\fn{It's of course true that the magnitude of the particle's acceleration is non-zero during the interval $\Delta t$---and, being Lorentz-invariant, this points to a genuine physical difference while the force is being applied---although, unlike Newtonian dynamics, in special relativity the non-zero magnitude of that acceleration is frame-dependent.}

Other situations illustrate the same point. Consider a world consisting of two otherwise identical particles moving in the same direction at different speeds relative to some inertial frame. Again, these particles occupy different dynamical states and thus possess different fundamental physical properties. If each were to collide with the same object, they would react differently. The 3-forces exerted in those collisions would be different. But it's not at all clear which properties to identify as accounting for their differing dynamical states, as the obvious candidates (noted above) either aren't fundamental or are shared between the particles. Again, what we'd like is some account of how these particles or states differ that only appeals to fundamental physical properties.\fn{In this instance we can see particularly clearly how the same issue arises for Newtonian dynamics, as the standard properties one might be inclined to appeal to aren't Galilean-invariant either and thus aren't candidates for being fundamental properties within the context of Newtonian theory (or don't differ between the particles). The motivations developed here thus apply equally to an analogous ontological picture of classical dynamics---although in that context there is no puzzle associated with energy and mass that the ontology helps to resolve.}

The postulation of a fundamental property of 4-momentum orientation provides an immediate and satisfying diagnosis of these situations. In the case of a particle subjected to a 3-force for a finite interval, the orientation of the particle's 4-momentum \emph{does} change. The orientations before and after the application of the force are distinct. Indeed, \emph{part} of what the relativistic force law says is that the dynamical effect of an interaction can be to change the 4-momentum orientation of a particle. Because I take 4-momentum orientation to be a fundamental physical property that partly defines a particle's dynamical state, changes in orientation provide an immediate explanation of how the dynamical state of the particle changes across $\Delta t$. In fact, it is the \emph{only} fundamental dynamical difference.\fn{Clearly, such changes in orientation are also associated with several coordinate-dependent effects, including changes in momentum and kinetic energy. This is how the 3-force component of a 4-force gives rise to frame-dependent changes in velocity and kinetic energy.} Similarly, what distinguishes the dynamical states of otherwise identical particles possessing different kinetic energies is that their 4-momenta point in different directions of spacetime, even though they might be moving in the same spatial direction. The ability to explain these distinctions is obscured if we look only within a frame, for a primitive notion of 4-momentum orientation is only a salient metaphysical option from a geometrical spacetime perspective.

\subsubsection{Interactions in Spacetime}

This is not the complete ontology. Turning to the dynamics of particle interactions, 4-forces themselves on this picture are fundamental, not 3-forces. Relativistic 3-forces are generally introduced as the analogues of Newtonian forces, so there's a temptation to assign them the same ontological status within each theory. But as has been emphasized throughout this paper, one principle reflected in special relativity is that all inertial frames are physically equivalent. A relativistic 3-force, as part of the spatial component of a 4-force, simply does not transform in a Lorentz-covariant way. Rather, 3-forces transform on the model of ordinary spatial velocity.\fn{See \citet[p.91]{wrindler91} for discussion.} Letting $S$ and $S^\prime$ be two inertial frames in standard configuration, $v$ the speed of $S^\prime$ relative to $S$ (along the x-axis), $\vec{u} \stackrel{S}{\rightarrow} (u_1,u_2,u_3) = (\frac{dx}{dt},\frac{dy}{dt},\frac{dz}{dt})$ the spatial velocity of a particle $p$ in frame $S$, and $\vec{f} \stackrel{S}{\rightarrow} (f_1,f_2,f_3)$ the S-components of the relevant 3-force applied to $p$, the components of the 3-force $\vec{f^\prime}$ experienced by $p$ in $S^\prime$ are given via the following transformations:
\begin{eqnarray}
  f_1^\prime & = & \frac{f_1 - v\frac{d (\gamma_u m)}{dt}}{(1-\frac{u_1v}{c^2})} \nonumber
\end{eqnarray}
\begin{eqnarray}
  f_2^\prime & = & \frac{f_2}{\gamma_v (1-\frac{u_1v}{c^2})} \nonumber
\end{eqnarray}
\begin{eqnarray}
  f_3^\prime & = & \frac{f_3}{\gamma_v (1-\frac{u_1v}{c^2})}. \nonumber
\end{eqnarray}
The transformation depends upon both the relative velocity between frames and (surprisingly) the velocity of the particle itself on which the force is acting. It is evident from these transformations that boosting from one frame to the next changes both the magnitude of the 3-force $p$ experiences and the angular difference between it and $p$'s 3-velocity $\vec{u}$. That is, if $p$ experiences a particular 3-force applied in a specific direction in a given frame $S$, the applied 3-force in a boosted frame $S^\prime$ will generally have both a different magnitude and a different direction.\fn{Only when the force is `pure' (discussed below) and the boost is in the direction of $\vec{u}$ will it be the case that $\vec{f} = \vec{f^\prime}$. See \citet[p.91]{wrindler91} for discussion.} There is no frame-independent fact about what 3-force is acting on a particle at any given event, and thus 3-forces are not the sorts of posits that ought to be taken as ontologically fundamental. (At least, one cannot take the 3-force as ontologically fundamental without privileging some particular inertial frame.) Indeed, there are situations in which a particle experiences no impressed 3-force in its rest frame but a non-zero 3-force in all other frames, each parallel to the direction in which the frame is boosted relative to the rest frame.\fn{\citet[p.92]{wrindler91} calls such forces `heatlike'.}

Instead, I propose that we understand interactions between particles by taking 4-forces as (or as representing something) fundamental. This is a natural proposal given the formal connection between 4-forces and 4-momenta changes and the fact that 4-momenta are here taken to encode the instantaneous dynamical states of particles. As with spatial forces in Newton's physics, 4-forces on this view are the fundamental things mediating interactions between particles and changing their dynamical states.

Like 4-momenta, 4-forces possess magnitudes and orientations in spacetime. We can think of 4-forces as pushes and pulls \emph{through spacetime}. Whereas the push or pull of a Newtonian force tries to change the way a body is moving through space, a 4-force tries to change the way a body is moving through spacetime.\fn{This view of Newtonian forces as pushes and pulls is emphasized in \citet{jwilson07}.}  That spacetime isn't isotropic means that the character of the push or pull exhibits itself in different ways depending on: the orientation of the 4-force; the orientation of the 4-momentum of the particle to which it's applied; and the frame in which it's being considered. For example, a 4-force applied orthogonally to a particle's 4-momentum and considered in the rest frame of the particle will manifest itself as a purely spatial push. In other frames it will also involve a `temporal push', in each case rotating the orientation of the particle's 4-momentum.

In summary, then, the fundamental ontology I postulate for special relativistic particle dynamics is one according to which there are two types of basic entities, particles and 4-forces, each of which possess two distinct fundamental properties---one represented by a scalar magnitude and one associated with an orientation in spacetime. (This is in addition to Minkowski spacetime and any properties that might come along with that.) Many of the entities and properties we typically associate with the dynamics of particles in special relativity are, on the proposal here, not fundamental.

\section{The Geometry of Composite Mass}

Returning to the puzzle of composite mass, recall that the issue there was that there seemed to be a variety of cases in which a physical process was occurring in a system even though none of the fundamental properties of the system's constituents seemed to be changing. I agree with Lange that no physical conversion occurs between mass and energy when, say, a gas is heated, and yet a real physical process \emph{is} occurring that is changing the system's mass. If the gas were enclosed in a box, it would be harder to move after being heated. What we would like is an explanation of how the mass of the system is constituted by fundamental physical properties of its underlying constituents---properties that \emph{are} changing as the gas is heated.

The trick is to recognize that velocity-dependent quantities are not the only features of the molecules that are changing in this process. Even if the magnitudes of the 4-momenta remain constant, the orientations of the 4-momenta are changing in response to whatever 4-forces are transmitting the heat. On the ontology advocated here, these are fundamental physical changes to the dynamical states of the constituent molecules. Recognizing that the 4-momentum of the total system is just the ordinary vector sum in Minkowski spacetime of its constituent 4-momenta, it follows that these changes in 4-momenta orientations at the constituent level give rise to the changing mass of the gas as a whole.\fn{As noted in footnote \ref{nofields}, the situation is more complicated for systems whose physical descriptions require field-theoretical considerations, for in those cases there isn't generally a well-defined total 4-momentum that can be associated with the system.} To see this, let's start with the diagram below for the (much simplified) case of a 2-body system:
\begin{center}
	\includegraphics[width=3in]{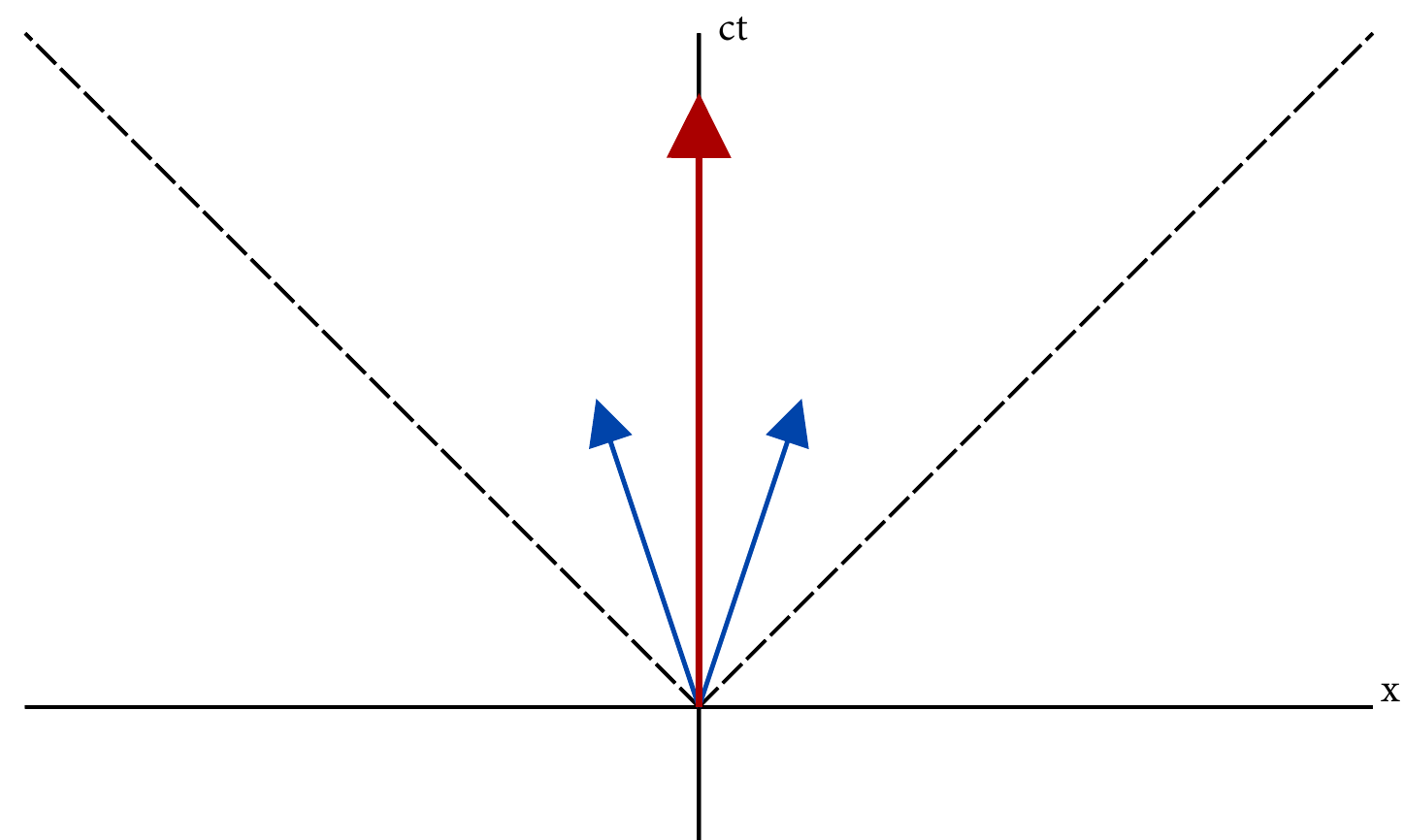}
\end{center}
For any gas, of course, there are vastly more component 4-momentum vectors, although they too add up to just a single net 4-momentum for the system. Notice that the magnitude of the net 4-momentum (in red) clearly depends on both the magnitudes \emph{and the orientations} of the constituent's 4-momenta.

Consider now what happens as the system is heated, as the orientations of the constituent 4-momenta change but not their magnitudes (masses). Due to the geometry of Minkowski spacetime all such equal-magnitude 4-vectors have endpoints that lie along the hyperbola shown in the following diagram:
\begin{center}
	\includegraphics[width=3in]{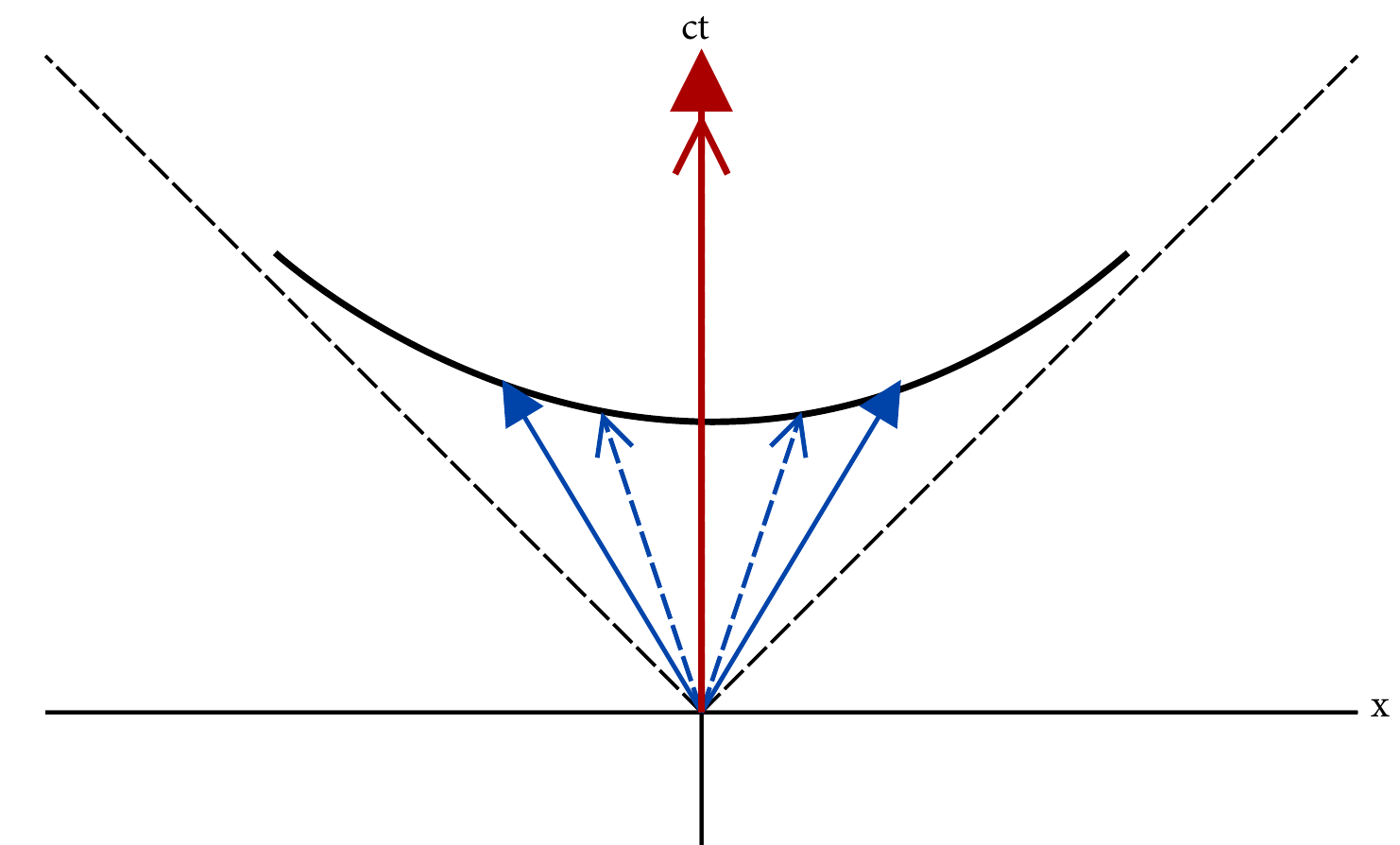}
\end{center}
The hyperbola itself is Lorentz invariant, although the inclinations at which the 4-vectors are drawn is frame-dependent. Here, the new 4-momenta (the solid vectors) are drawn in the original rest frame of the system. Notice that the 4-momentum of the system does not remain constant during the heating process. It retains its orientation in spacetime as the orientations of its constituent 4-momenta change, but---unlike its constituents---its magnitude also changes. As the constituent particles move faster in the original frame---as their 4-momenta rotate their orientations---the magnitude of the system's 4-momentum gets larger. So not only can the composite mass of the system be understood as constituted by fundamental physical properties of its molecular constituents (namely, the physical properties encoded in the constituents' 4-momentum vectors) but we can understand what genuine physical changes are occurring at the molecular level when the gas is heated that result in changing the mass of the gas as a whole. Unlike changes in velocity and kinetic energy, these are fundamental physical changes, and as such provide a solution to the original puzzle of composite mass.

Although Lange is right that the mass of an $n$-body system can be expressed as $$M = (m_1 + \dots + m_n) + \frac{1}{c^2}(T_1 + \dots + T_n),$$ the dependence of system mass on constituent energies is a coordinate-dependent manifestation of the more fundamental dependence of system mass on 4-momenta magnitudes and orientations. Lange's expression is an artifact of how the magnitude of the system's total 4-momentum decomposes into its constituents' 4-momentum components in the center-of-momentum frame. It does not reflect the genuine physical properties that actually determine the mass of the gas. Those are the constituent 4-momentum magnitudes and orientations, a fact which (again) is obscured unless one adopts the 4-dimensional ontology of particle dynamics proposed here.

\section{Inellastic Collisions at the Level of `Principle'}

Turning to inelastic collisions, the 4-Momentum Collision Principle tells us how the total 4-momentum of an isolated system constrains the way in which the 4-momenta magnitudes and orientations of its constituents change when they undergo collision. Initially (i.e., pre-collision), the net 4-momenta of each ball are oriented in different directions, and that's why the two colliding bodies have a total mass that is more than the sum of the two bodies themselves. (That is, the magnitude of the total 4-momentum vector is greater than the sum of the magnitudes of the two component 4-momentum vectors, as was the case with the gas.) When the two bodies collide inelastically, each changes its 4-momentum orientation such that both come to be oriented in the same direction. Because the 2-body system is isolated, however, the total 4-momentum before and after collision must remain the same. This means that when the colliding bodies change their 4-momenta upon impact, their new 4-momentum vectors must each be oriented in the same direction as the total system 4-momentum prior to impact \emph{and} must preserve the magnitude of the total system 4-momentum prior to impact. But this can only happen if, when the orientations change upon impact, their magnitudes \emph{increase} as well. This means that the mass of each body goes up upon impact, which is why if the resulting body post-collision were cut in half, each half would exhibit more inertial resistance than it did pre-collision. Of course, this consequence is an explanation at the `principle' level---it doesn't tell us anything about the underlying dynamical mechanism in virtue of which this change might be brought about. That point will be addressed in the next section and lies at the heart of what is truly novel about special relativistic particle dynamics.

\section{Elementary Conversion and $E_0=mc^2$}

Let us (finally) address perhaps the most puzzling case of all---apparent cases of genuine conversion between mass and energy in the context of elementary particle collisions. What might really be going on in these reactions, at least when viewed through the lens of special relativistic particle dynamics? Recall that in electron and positron creation collisions the incoming photons are massless but the outgoing particles have mass. How do we reconcile what's occurring with the principles of special relativity, without appealing to kinetic energy as a fundamental physical property?

If we consider the system of colliding photons that initiate the collision reaction, their 4-momenta (in green) are represented in the following center-of-momentum frame diagram:
\begin{center}
	\includegraphics[width=3in]{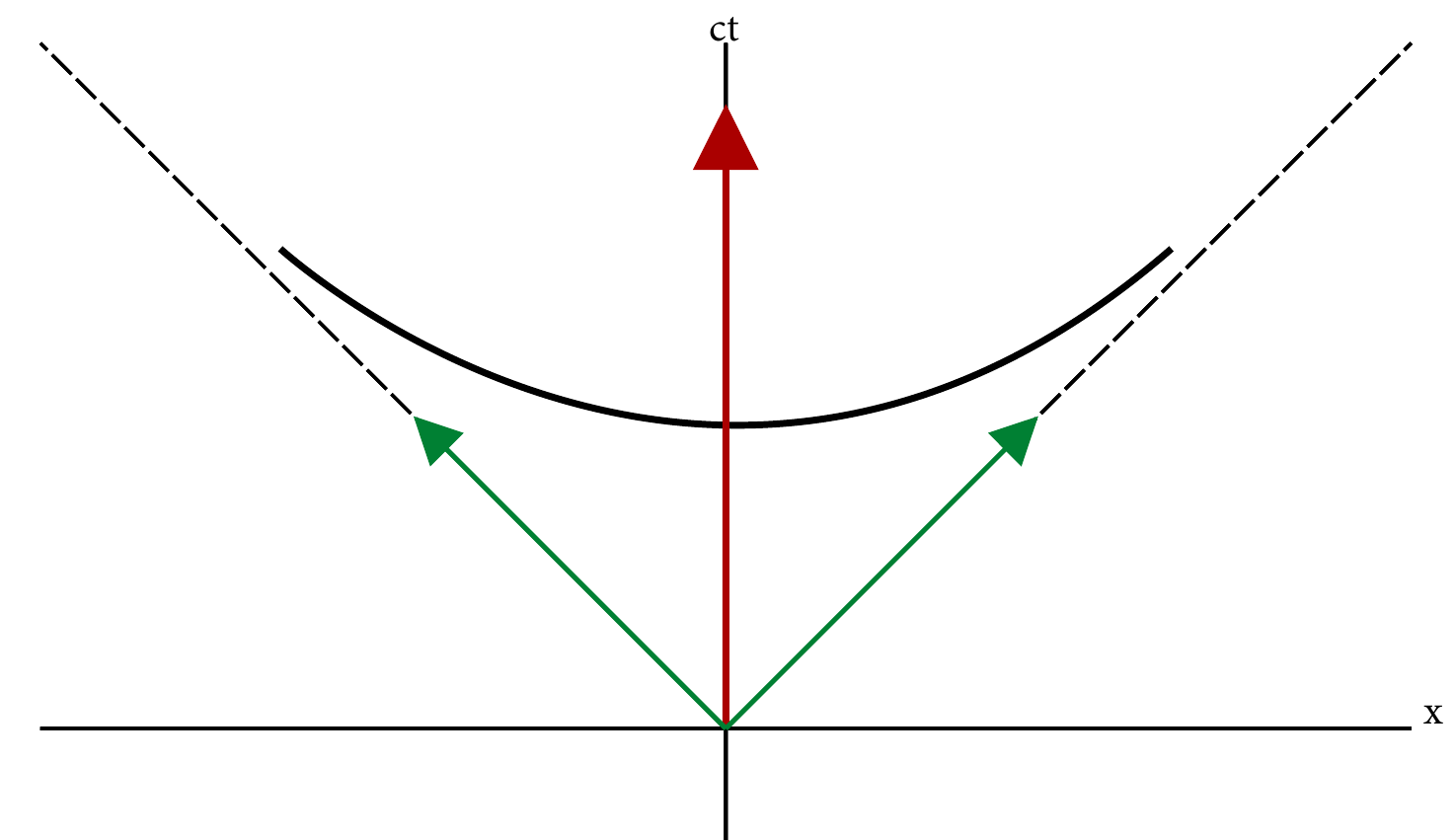}
\end{center}
The 2-photon system as a whole clearly possesses a non-zero 4-momentum magnitude (and hence a non-zero mass), even though neither constituent does. According to the 4-Momentum Collision Principle this total 4-momentum gets preserved in the interaction, as in the following diagram:
\begin{center}
	\includegraphics[width=3in]{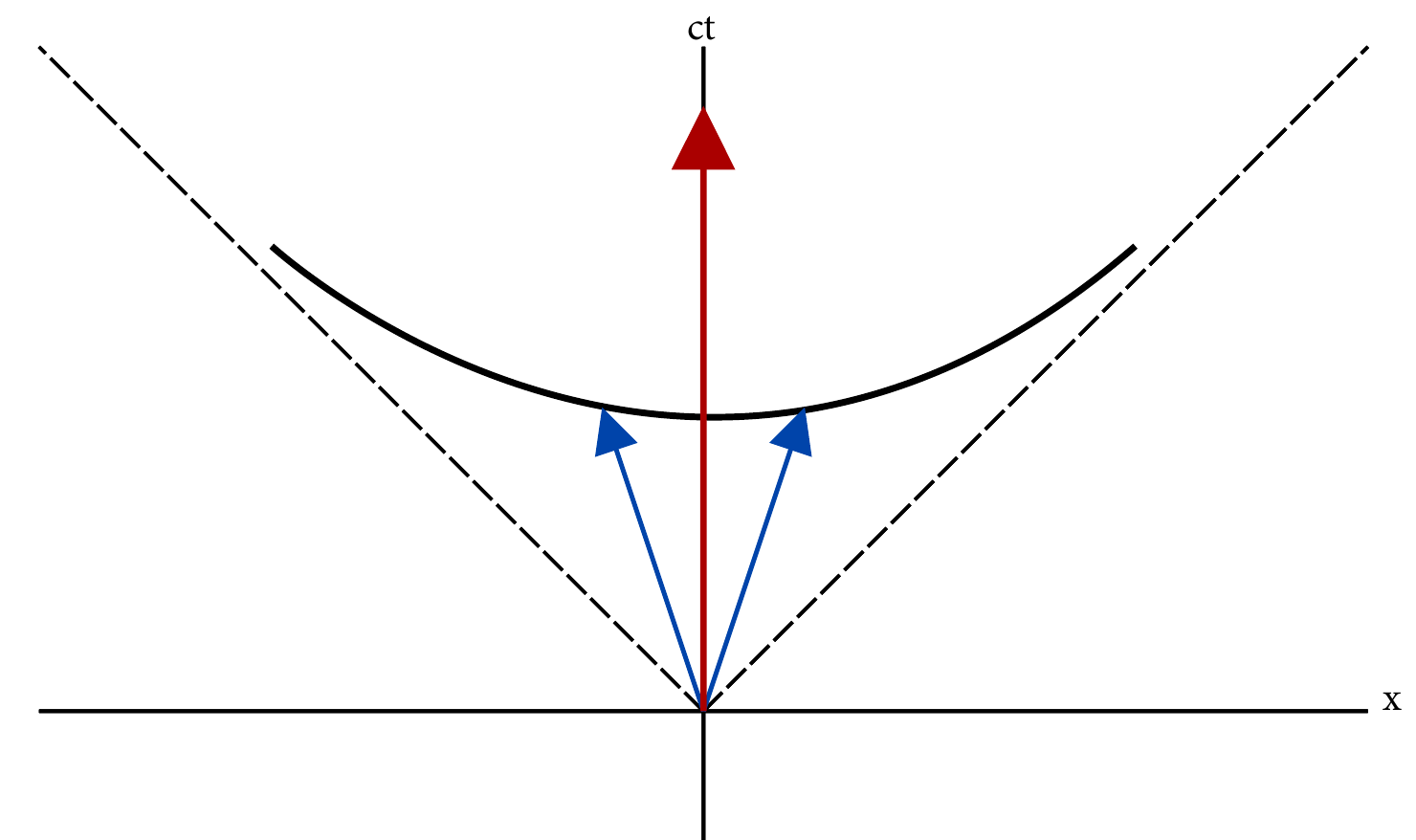}
\end{center}
As with the inelastic collision case, we can straightforwardly deduce from the 4-Momentum Collision Principle \emph{that} the output particles will exhibit different inertial resistances than the inputs. For the inertial resistances of the output particles are determined by their 4-momentum magnitudes, and those in turn are determined by the 4-momentum magnitudes \emph{and orientations} of the input particles in accordance with the 4-Momentum Collision Principle. There is a genuine transformation of objective physical properties in this reaction, but it's not a conversion between mass and energy. What changes in the course of the reaction are simply the magnitudes and orientations of the constituent 4-momenta. Mass is not created \emph{out of something else}, as the original puzzle seemed to suggest. The transformation in this reaction is from one set of 4-momenta magnitudes and orientations to another. Mass (as 4-momentum magnitude) simply arises in virtue of these changes.

Still, what we want---what is needed to complete the picture in a satisfying way---is some account of the dynamics that underpins this transformation. Can a dynamical story be given, in terms of 4-momentum magnitudes and orientations, according to which constituent masses `emerge' in this reaction? This is something that my interpretation of special relativistic particle ontology ought to provide. (Again, we shouldn't expect special relativity to offer an explanation of this \emph{specific} collision reaction, which is quantum in nature.)

The answer, I think, turns on the nature and reality of 4-forces. When the incoming constituents collide, they experience 4-forces that change their 4-momenta. Most 4-forces are such that they change only the orientation but not the magnitude of a particle's 4-momentum, as in the case of the heated gas. A 4-force of this sort has (in a frame) the following components: $$F^\mu = \gamma_u (cm\frac{d\gamma_u}{dt},\vec{f}),$$ where the associated relativistic 3-force $$\mathbf{f} = m\frac{d\gamma_u}{dt}\vec{u} + \gamma_um\vec{a}$$ is called a `pure' force. The electromagnetic force on a charged body is the canonical example of a pure 3-force. As \citet[p.92]{wrindler91} shows, a necessary and sufficient condition for a 3-force to be pure is that its associated 4-force $\mathbf{F}$ always satisfies $\mathbf{U} \cdot \mathbf{F} = 0$. As a consequence, if a 4-force always acts orthogonally in Minkowski spacetime to a particles 4-velocity (4-momentum), then the magnitude of the 4-momentum will be preserved. However, a second possibility is also built into the dynamical structure of $\mathbf{F} = \frac{d\mathbf{P}}{d\tau}$ -- namely, that a 4-force might act in such a way as to change the \emph{magnitude} of a particle's 4-momentum. Such forces can be expressed in a frame as $$F^\mu = \gamma_u (cm\frac{d\gamma_u}{dt} + c\gamma_u\frac{dm}{dt},\vec{f}),$$ and their associated 3-forces $$\mathbf{f} = (m\frac{d\gamma_u}{dt} + \gamma_u\frac{dm}{dt})\vec{u} + \gamma_um\vec{a}$$ are said to be `impure'. By their dynamical action alone, 4-forces of this sort are capable of changing the mass of a particle.\fn{A special case of such forces are what \citet[p.92]{wrindler91} dubs `heatlike' forces, the action of which only changes a particles 4-momentum magnitude and not its orientation.} Naturally, a necessary and sufficient condition for a 3-force to be impure is that $\mathbf{U} \cdot \mathbf{F} \neq 0$ hold for the associated 4-force. \citet[p.92]{wrindler91} shows using this condition that any 4-force derivable from a 4-scalar potential $\Phi$ via an equation of the form $F_\mu = \frac{\partial\Phi}{\partial x^\mu}$, such as the scalar meson theory of the nucleus, must be impure.

Return now to the electron-positron creation reaction and to a diagram of possible timelike oriented outputs:
\begin{center}
	\includegraphics[width=3in]{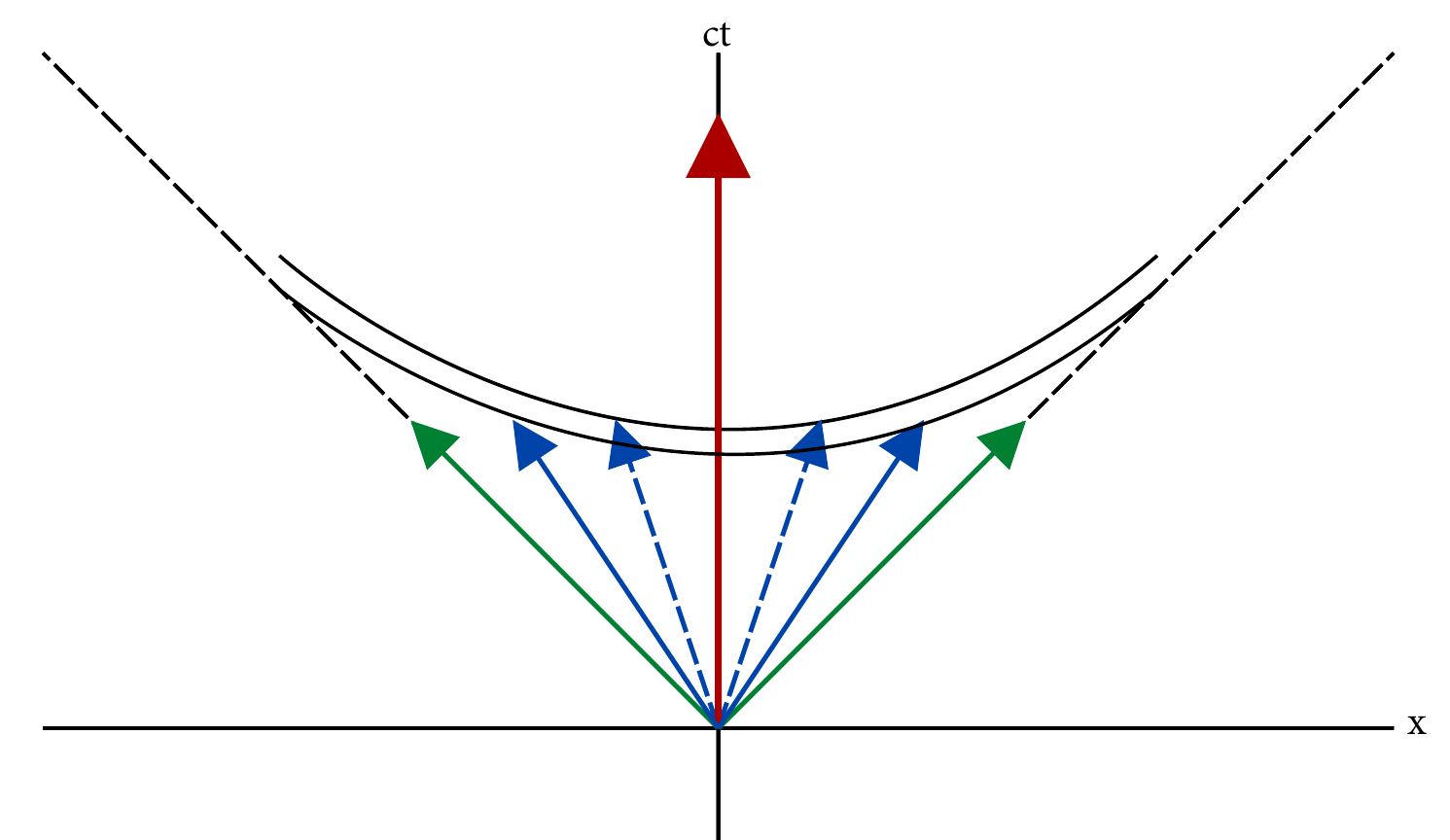}
\end{center}
As one can see from the diagram, if the outputs are not null-oriented the 4-forces applied must be such that $\mathbf{U} \cdot \mathbf{F} \neq 0$. (Recall that for a null 4-vector $\mathbf{U}$ in Minkowski spacetime, $\mathbf{U} \cdot \mathbf{F} = 0$ iff $\mathbf{F}$ is \emph{parallel} to $\mathbf{U}$.) So if the 4-forces exerted are such that they change the \emph{orientations} of null 4-momenta, they must also change their \emph{magnitudes}. This holds because the applied forces in the reaction (whatever they are exactly) must be impure, and it drops right out of the dynamical equations.\fn{A similar story can be given for the case of inelastic collisions, as discussed in the preceding section. Evidently the 4-forces at work in such a collision must be impure so as to change the magnitudes of the incoming particles' 4-momenta.}

Mass, then, as the property of inertial resistance, can be created and destroyed in special relativistic dynamics, and it is changes in 4-momenta magnitudes brought about by impure 4-forces that provide the dynamical mechanisms for these changes.\fn{Many textbooks are blithely indifferent to the possibility of impure forces. See, e.g., \citet[p.215]{afrench68} and \citet[p.192]{jfreund08}, where $\frac{dm}{dt} = 0$ is assumed without any comment whatsoever.} \emph{This} is what is truly dynamically novel about special relativistic particle mechanics: mass can be created and destroyed by 4-forces. Now, as it happens, the application of a 4-force (in general) \emph{also} changes the energy of a particle, although the \emph{rate} of that change---just like the magnitude and the spatial direction of the 3-force---is frame-dependent. This misleads us into thinking that in some cases there is a conversion between energy and mass. But there is no such conversion. Instead, what happens is that impure 4-forces change the 4-momentum magnitudes of the particles on which they act while also changing the energies of those particles. \emph{What $E_0=mc^2$ tells us is how changes in particle mass brought about by 4-forces are correlated with changes in energy brought about by those same 4-forces.}\fn{In this sense, one really ought to stick to writing the equation as $\Delta E_0 = \Delta mc^2$, which is the form that actually gets used in physical practice.} But the change in mass is a fundamental physical change, whereas the change in energy is a change in a non-fundamental or derivative property. That there is a precise and fixed correspondence between these two changes is an important empirical fact, arising from the action of 4-forces and facts about their frame-dependent decompositions. In this sense $E=mc^2$ \emph{does} encode a profound and novel dynamical discovery, but that discovery can't be read off of the equation itself and it holds in virtue of the nature and existence of 4-forces.\fn{The ontology proposed here deals with the apparent reality of energy released in, say, a nuclear explosion by saying that the energy released is a frame-dependent effect of the 4-momentum magnitudes changing on account of the applied (impure) 4-forces.} $E_0=mc^2$ does not mean what it is often taken to mean, or what Einstein took it to mean.

\section{Conclusion}

Although often taken to be the central insight of special relativistic particle dynamics, the alleged equivalence between energy and mass expressed in Einstein's famous equation $E_0=mc^2$ remains both controversial amongst physicists and conceptually problematic, as it contravenes a central principle regarding the content and interpretation of special relativity---namely, that all inertial frames are physically equivalent. Nevertheless, the apparent inter-convertability of energy and mass seems to have \emph{prima facie} experimental support. In this paper I have proposed a specific account of the fundamental ontology of special relativistic particle dynamics that resolves this tension. In the process, I have argued that the core dynamical insight of special relativity lies not in any relationship between energy and mass, but rather in the recognition that the frame-independent application of 4-forces to particles can in some cases actually bring about changes in particle mass. That changes in a primitive property (mass) are correlated with changes in a derivative property (energy) is a byproduct of that fundamental dynamical fact.

\pagebreak
\renewcommand{\baselinestretch}{1}\small\normalsize
\bibliographystyle{phil}
\bibliography{philreferences}

\end{document}